\documentclass[reprint, amsmath,amssymb,aip, sd]{revtex4-1}

\usepackage{graphicx}
\usepackage{rotating}
\usepackage{amssymb}
\usepackage{mathptmx}
\usepackage{lipsum}
\usepackage{float}

\begin{document}

\title{Suppression of radiation loss in high kinetic inductance superconducting co-planar waveguides}
\author{S. H\"{a}hnle}
 \altaffiliation{s.haehnle@sron.nl}
\affiliation{SRON - Netherlands Institute for Space Research, Sorbonnelaan 2, 3584 CA Utrecht, The Netherlands}
\affiliation{Department of Microelectronics, Faculty of Electrical Engineering, Mathematics and Computer Science (EEMCS), Delft University of Technology, Mekelweg 4, 2628 CD Delft, The Netherlands}

\author{N. v. Marrewijk}
\affiliation{Department of Microelectronics, Faculty of Electrical Engineering, Mathematics and Computer Science (EEMCS), Delft University of Technology, Mekelweg 4, 2628 CD Delft, The Netherlands}

\author{A. Endo}
\affiliation{Department of Microelectronics, Faculty of Electrical Engineering, Mathematics and Computer Science (EEMCS), Delft University of Technology, Mekelweg 4, 2628 CD Delft, The Netherlands}
\affiliation{Kavli Institute of NanoScience, Faculty of Applied Sciences, Delft University of Technology, Delft, The Netherlands.}

\author{K. Karatsu}
\affiliation{SRON - Netherlands Institute for Space Research, Sorbonnelaan 2, 3584 CA Utrecht, The Netherlands}
\affiliation{Department of Microelectronics, Faculty of Electrical Engineering, Mathematics and Computer Science (EEMCS), Delft University of Technology, Mekelweg 4, 2628 CD Delft, The Netherlands}

\author{D. J. Thoen}
\affiliation{Department of Microelectronics, Faculty of Electrical Engineering, Mathematics and Computer Science (EEMCS), Delft University of Technology, Mekelweg 4, 2628 CD Delft, The Netherlands}

\author{V. Murugesan}
\affiliation{SRON - Netherlands Institute for Space Research, Sorbonnelaan 2, 3584 CA Utrecht, The Netherlands}

\author{J. J. A. Baselmans}
\affiliation{SRON - Netherlands Institute for Space Research, Sorbonnelaan 2, 3584 CA Utrecht, The Netherlands}
\affiliation{Department of Microelectronics, Faculty of Electrical Engineering, Mathematics and Computer Science (EEMCS), Delft University of Technology, Mekelweg 4, 2628 CD Delft, The Netherlands}

\begin{abstract}
We present a lab-on-chip technique to measure  the very low losses in superconducting transmission lines at (sub-) mm wavelengths. The chips consist of a 100 nm thick NbTiN Coplanar Waveguide (CPW) Fabry-P{\'e}rot (FP) resonator, coupled on one side to an antenna and on the other side to a Microwave Kinetic Inductance detector. Using a single frequency radiation source allows us to measure the frequency response of the FP around 350 GHz and deduce its losses.  We show that the loss is dominated by radiation loss inside the CPW line that forms the FP and that it decreases with decreasing line width and increasing kinetic inductance as expected.  The results can be quantitatively understood using SONNET simulations. The lowest loss is observed for a CPW with a total width of $6\ \mathrm{\mu m}$ and corresponds to a Q-factor of $\approx15,000$.
\end{abstract}

\maketitle

Superconducting transmission lines, such as co-planar waveguides (CPWs) or microstrips, are increasingly prevalent for cryogenic high-frequency applications upwards of 100$\ $GHz, such as on-chip spectrometers\cite{DeshimaNature, Cataldo2018, Superspec}, phased array antennas\cite{Ade_2015} and kinetic inductance parametric amplifiers \cite{ho_eom_wideband_2012}. These applications require ultra low-loss transmission lines with a loss tangent of $\tan\delta\lesssim10^{-3}$ and lengths upwards of 100$\lambda$, either as an integral part of the circuit in kinetic inductance parametric amplifiers or phased array antennas, or as a connecting element in on-chip spectrometers.  Microstrip losses in this frequency range down to $\tan\delta=2\times10^3$ have been measured previously\cite{Gao2009}. Here, we focus on losses in CPW.CPW lines have an advantage over microstrip lines in that they do not require a deposited dielectric, which is a source of loss, decoherence and noise. However, CPWs are open structures and can radiate power, which is a source of loss and increases cross coupling to neighboring lines. The dominant radiation loss mechanism is the so-called leaky mode, which is present if the phase velocity in the line exceeds the phase velocity in the substrate. For microwave applications, this can be controlled by reducing the line width, but this becomes increasingly impractical at mm- and sub-mm wavelengths. In superconducting lines, the phase velocity is reduced due to kinetic inductance, which in principle allows to create a line with a phase velocity below the substrate phase velocity, thereby eliminating the leaky mode radiation and creating ultra low-loss transmission lines at frequencies exceeding hundreds of GHz. Dielectric losses in microstrips at frequencies up to 100 GHz have been measured previously\cite{Gao2009},   In this paper, we demonstrate lab-on-chip loss measurements of superconducting NbTiN CPW Fabry-P{\'e}rot resonators around 350 GHz. We show that the radiation loss can be reduced and even virtually eliminated by reducing the phase velocity, which is accomplished by narrowing the CPW line to a total width of $\lesssim6\ \mu$m.


\begin{figure}
\includegraphics{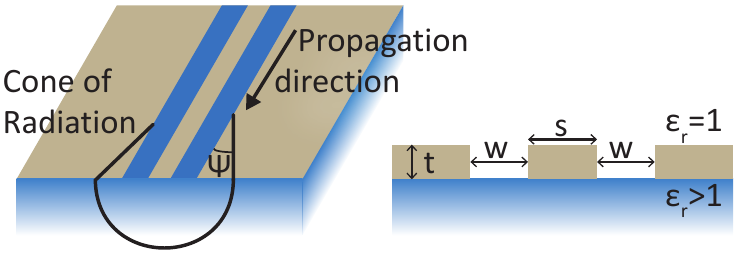}
\caption{a) CPW geometry. b) Cone of radiation emitted along the propagation direction of the CPW mode, with radiation angle $\Psi$.}
\label{fig:cpw}
\end{figure}

\begin{figure*}[ht]
	\includegraphics{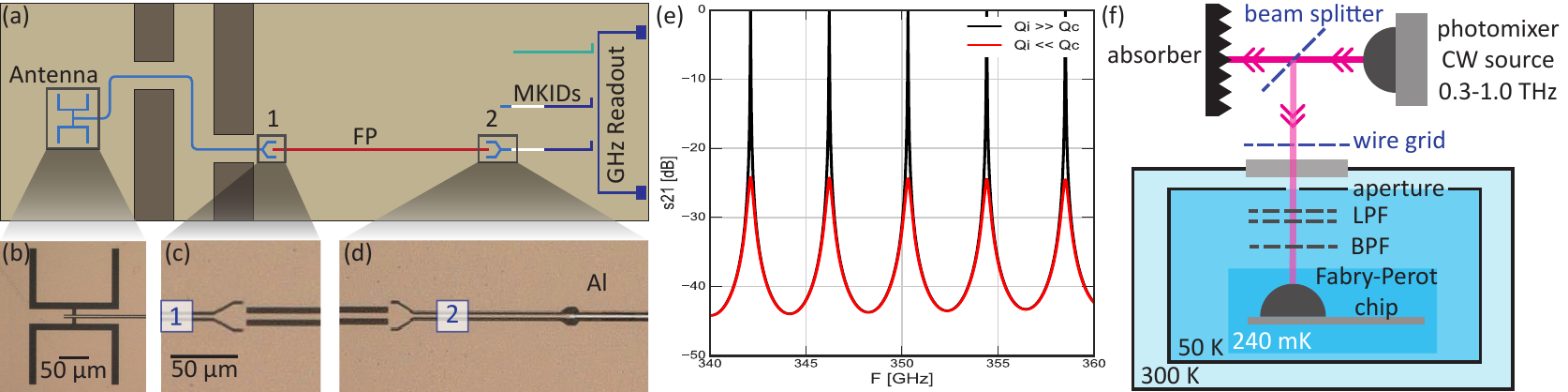}
	\caption{a) Chip schematic.  b) Picture of antenna. c) Picture of first coupler. d) Picture of second coupler, including the transition to the Aluminum section of the MKID.  e) Simulated Fabry-P{\'e}rot transmission. f) Experimental setup schematic. The filterstack consists of Low-pass filters (LPF) and a bandpass filter (BPF). The aperture plane is at the 50K window and a polarizing wire grid is located outside the cryostat.}
	\label{fig:FP} 
\end{figure*}

The effective dielectric constant of a transmission line using a perfect electric conductor (PEC) is given by 
\begin{equation} \label{eq:eeff}
\epsilon_{eff}=c^2LC
\end{equation}
where $c$ is the speed of light and $L$ and $C$ are the transmission line inductance and capacitance per unit length respectively. In a CPW as shown in Fig. \ref{fig:cpw}, this can be approximated by
\begin{equation}
\epsilon_{eff}\approx\frac{\epsilon_r+1}{2}
\end{equation}
with the dielectric constant of the substrate $\epsilon_r$. The phase velocity $v_{ph} = \frac{c}{\sqrt{\epsilon_{eff}}}$ in the guided CPW mode is therefore faster than in the substrate. This creates a shockwave in the substrate, leading to a radiation cone characterized by the radiation angle $\Psi$ (see Fig. \ref{fig:cpw}). The frequency dependent loss factor $\alpha$ at high frequencies due to this shockwave has been derived by Frankel et al. \cite{Frankel92}  from the electric and magnetic field distributions in the dielectric materials due to the current distribution in a PEC as
\begin{equation} \label{eq:frankel}
\alpha_{rad} = \left(\frac{\pi}{2}\right)^5 2\left(\frac{\left(1-\cos^2(\Psi)\right)^2}{\cos(\Psi)}\right)\frac{(s+2w)^2\epsilon_r^{3/2}}{c^3K(\sqrt{1-k^2})K(k)}f^3
\end{equation}
where $s$ and $w$ are the CPW line and slot width, $k = s/(s+2w)$ and $K(k)$ is the complete elliptical integral of the first kind. 
It can be seen in Eqn.\ref{eq:frankel}, that the magnitude of radiation loss is strongly dependent on $\Psi$ which is given by the discrepancy of the dielectric constants
\begin{equation} \label{eq:Psi}
\cos(\Psi) = \frac{\sqrt{\epsilon_{eff}(f)}}{\sqrt{\epsilon_r}}.
\end{equation}

For a PEC CPW, this ratio is only dependent on the substrate and independent of the conductor properties. However, in a superconducting CPW, the kinetic inductance per unit length $L_k$ due to the inertia of Cooper pairs needs to be taken into account, changing Eqn. \ref{eq:eeff} to
\begin{equation} \label{eq:eeff_sc}
\epsilon_{eff}=c^2(L_g+L_k)C
\end{equation}
where the transmission line inductance is the sum of its kinetic inductance and geometric inductance $L_g$. 

Conceptually, using a CPW with high $L_k$ leads to a suppression of the radiation loss, as the radiative angle $\Psi$ is reduced. If $L_k$ is sufficiently large to obtain $\epsilon_{eff}\geq\epsilon_r$, the radiative shockwave does not form as the phase velocity of the CPW line is slower than in the substrate, resulting in theoretically zero radiation loss. The kinetic inductance Lk increases with the film normal state sheet resistance, a reduced film thickness (in the regime of thin films compared to the penetration depth) and with reducing linewidth. A CPW of a 100 nm NbTiN film of in total $6\ \mathrm{\mu m}$ wide will fulfill the condition that $\epsilon_{eff} > \epsilon_r$ (see supplementary material). Another method is to use a CPW fabricated on a vanishingly thin dielectric membrane, which can be approximated as a free standing CPW and therefore does not radiate.

Measuring the radiation loss of a superconducting CPW at sub-mm wavelengths requires a highly sensitive device, capable of measuring a loss tangent $\tan\delta<10^{-3}$. For this purpose we design a chip with a Fabry-P{\'e}rot resonator at its core, as shown in the schematic of \ref{fig:FP}a). A similar device as been used by G{\"o}ppl et al.\cite{Goeppl08} at microwave frequencies.


The Fabry-P{\'e}rot (FP) resonator is a single CPW line terminated by two identical couplers on either end, with the resonance condition  
\begin{equation} \label{eq:Fn}
	F_n = n\frac{c}{2L_{FP}\sqrt{\epsilon_{eff}}} 
\end{equation}
where $\epsilon_{eff}$ is the effective dielectric constant of the CPW, $L_{FP}$ is the resonator length and $n$ is the mode number. Transmission through the resonator can be described as a series of Lorentzian peaks, where each peak has a loaded Quality factor $Q_L$ given by the resonance frequency and FWHM (full width at half maximum)
\begin{equation}
\label{eq: Ql}
	Q_L = \frac{F_n}{\mathrm{FWHM}_n}.
\end{equation}
The loaded Q-factor is a measure of the power loss per cycle which can be separated in its two primary components:
\begin{equation}\label{eq:allQ}
\frac{1}{Q_L} = \frac{1}{Q_c} + \frac{1}{Q_i}
\end{equation}

First, $Q_c$ is the power leakage through the two couplers 
\begin{equation}\label{eq:Qc}
	Q_c(n) = \frac{n\pi}{\left|S_{2'1'}\right|^2}
\end{equation}
where $\left|S_{2'1'}\right|^2$ is the transmission through a single coupler with ports $1'$ and $2'$ (see the supplementary material). Second, the internal losses described by $Q_i$ which is defined as  
\begin{equation}
	Q_i = \frac{\beta}{2\alpha}
\end{equation}
with the propagation constant $\beta = 2\pi/\lambda$ and the loss factor $\alpha$, where $[\alpha] = \mathrm{Np/m}$. The loss inside the resonator is given by the combination of ohmic loss ($Q_{i,ohm}$), dielectric loss ($Q_{i,diel}$) and radiation loss of the CPW ($Q_{i,rad}$), as well as radiation loss at the coupler ($Q_{i,coup}$):
\begin{equation}
\frac{1}{Q_i}= \frac{1}{Q_{i,ohm}} + \frac{1}{Q_{i,diel}} + \frac{1}{Q_{i,rad}} + \frac{1}{Q_{i,coup}} 
\end{equation}
Since $Q_L$ is the measured variable, a precise measurement of $Q_i$ requires exact knowledge of $Q_c$, which is experimentally difficult due to fabrication constraints. Therefore, measurements in the internal loss dominated regime of $Q_c>Q_i$ are preferred, since then $Q_L\approx Q_i$ (see Eqn.\ref{eq:allQ}). However $Q_c$ cannot be increased arbitrarily, as this will reduce the Lorentzian peak height according to   
\begin{equation}
\label{eq:s21max}
	|S_{21}|_{max}= \frac{Q_L}{Q_c},
\end{equation} 
as shown in Fig \ref{fig:FP}e). Additionally, we use in the experiments a source with limited frequency resolution, limiting the design range of $Q_c$ as well. Taking these considerations into account, all chips discussed in this paper are designed in Sonnet\cite{manual2008sonnet} to have $Q_c^{design} = 2.7\times10^4$ at $350\ \mathrm{GHz}$ (see the supplementary material).
The center frequency of $350\ \mathrm{GHz}$ is chosen based on the available experimental setup.

In order to measure the CPW radiation loss dependency on $\epsilon_{eff}$ in eqs.(\ref{eq:frankel}) and (\ref{eq:Psi}), four chips are designed with varying linewidths $w$ and slotwidths $s$ of the Fabry-P{\'e}rot lines as given in Table \ref{tab:geometry}. All chips are fabricated on a single $350\ \mathrm{\mu m}$ thick Sapphire wafer, ensuring common film properties across the chips. The $100\ \mathrm{nm}$ NbTiN film is deposited directly on the Sapphire using reactive sputtering of a NbTi target in a Nitrogen-Argon atmosphere\cite{Thoen_NbTiN}.  Details on the fabrication can be found in Endo et al. \cite{EndoJatis}, which follows the same route as this paper.

The measured line geometry is determined via SEM (Scanning Electron Microscope) inspection and deviates slightly due to overetch in the fabrication process. Using the surface inductance of $L_s = 1.03\ \mathrm{pH}$ at $350\ \mathrm{GHz}$ calculated from the measured film parameters ($T_c = 14.7\ \mathrm{K}, \rho_N = 102\ \mathrm{\mu\Omega cm}$), and the known parameters of the C-plane Sapphire substrate ($\epsilon_{r}^C=11.5$; $\epsilon_{r}^{AB}=9.3$ ), we obtain the measured values of $\epsilon_{eff}$ at $350\ \mathrm{GHz}$ given in Table \ref{tab:geometry}. NbTiN parameters are measured on a test sample close to the FP resonators to eliminate effects of spatial variations in the NbTiN properties\cite{Thoen_NbTiN}. The resonator length is $L_{FP} = 10\ \mathrm{mm}$, corresponding to mode numbers in the range of 60-90 for the four chips.

\begin{table}[b]
	\begin{tabular}{l|c|c|c|c}
		\textbf{} & \multicolumn{1}{l|}{$\mathrm{w_{d}=s_{d}\ [\mu m]}$} & \multicolumn{1}{l|}{$s_{meas}\ [\mu m]$} & \multicolumn{1}{l|}{$w_{meas}\ [\mu m]$} & \multicolumn{1}{l}{$\epsilon_{eff}$} \\ \hline
		Chip I    & 2                                                              & 1.95                                     & 2.15           & 13.1                         \\
		Chip II   & 3                                                              & 2.95                                     & 3.15           & 10.9                        \\
		Chip III  & 4                                                              & 3.95                                     & 4.15           & 9.5                        \\
		Chip IV   & 5                                                              & 4.95                                     & 5.15           & 9.0                       
	\end{tabular}
	\caption{Designed and measured slot width $w$ and line width $s$ and resulting $\epsilon_{eff}$ for each chips Fabry-P{\'e}rot resonator.}
	\label{tab:geometry}
\end{table}

A first estimate of the radiation loss, naively using Eqn.\ref{eq:eeff_sc} in Eqn.\ref{eq:frankel} to account for the kinetic inductance, ranges from $Q_i = 5.6\times10^3$ for the $5\ \mathrm{\mu m}$ line to $Q_i = 5.4\times10^6$ for the $3\ \mathrm{\mu m}$ line; for the $2\ \mathrm{\mu m}$ line, the equation diverges. 

Using Mattis-Bardeen theory\cite{MattisBardeen}, we can estimate the ohmic losses to be multiple orders of magnitude higher than the stated loss, which means that radiation loss dominates for $w = s > 2 \mathrm{\mu m}$. It has been shown previously, that highly disordered superconductors start to deviate from Mattis-Bardeen theory \cite{Driessen2012_PRL} for high frequencies ($f>0.3\Delta$) and high normal-state resistivity ($\rho_N>100 \mathrm{\mu\Omega cm}$). However, both the frequency range of this experiment and the NbTiN film resistivity are at the lower limit and only a minimal deviation is expected.

In order to drive the FP resonator, one coupler (port 1) is connected via a CPW with $w=2\ \mathrm{\mu m}$ and $s=2\ \mathrm{\mu m}$ to a double-slot antenna, centered at $350\ \mathrm{GHz}$. The other coupler (port 2) is directly attached to the shorted end of a Microwave Kinetic Inductance Detector (MKID), which is a $\lambda/4$ resonator with $F_{res}\approx6.5\ \mathrm{GHz}$ based on the hybrid CPW design introduced by Janssen et al.\cite{Janssen13}. In the MKID, a $1.5\ \mathrm{mm}$ long, narrow hybrid CPW with a NbTiN ground plane and an Al ($\Delta_{Al}\approx 90 \mathrm{GHz}$ $T_c=1.28\ K$) center line follows directly after a NbTiN coupler section as shown in Fig. \ref{fig:FP}d). Incoming THz radiation is absorbed in the low bandgap Al line, thereby creating quasiparticles which changes the kinetic inductance of the film. This causes a frequency shift of the MKID resonator which is read out with the SPACEKIDs microwave readout \cite{SpacekidsMUX}.

Additional MKIDs, which are not coupled to the Fabry-P{\'e}rot and hereafter referred to as blind MKIDs, are placed on the chip as reference detectors. A microwave resonator with the same CPW geometry as the FP resonator is also added (green line in Fig. \ref{fig:FP}a)). Sampling the full FP transmission requires a measurement with a dynamic range of $\approx 50\ \mathrm{dB}$ (see Fig. \ref{fig:FP}e)). In order to reduce stray light reaching the MKIDs, the copper holder in which the chip is placed contains a labyrinth structure as indicated in Fig. \ref{fig:FP}a), separating the chips exposed antenna section from the dark Fabry-P{\'e}rot section. Additionally, a low-Tc backside layer of beta-Ta is deposited on the chip backside and acts as a stray light absorber\cite{Yates2017}. 



In the experiment we mount an $8\ \mathrm{mm}$ Si lens on the chip backside, centered on the antenna, and place both in the Cu sample holder. This is placed on the cold stage of a He-3/He-4 sorption cooler\cite{HaehnleCryo}, as shown in Fig. \ref{fig:FP}f), operating at $T\approx250 \ \mathrm{mK}$. A commercial photomixer continuous wave (CW) source\cite{Toptica} is positioned at room temperature and coupled into the cryostat via a beamsplitter to reduce the incoming power and avoid saturation of the MKIDs. The source emits a linear polarized, single frequency signal which is tunable in the range of $0.1...1.2\ \mathrm{THz}$ with a minimum step size of $\sim10\ \mathrm{MHz}$ and an absolute frequency accuracy of $<2\ \mathrm{GHz}$. A band pass filter stack centered at $F_c = 346\ \mathrm{GHz}$ and $>20\ \mathrm{dB}$ out of band suppression is located in the cryostat with a polarizing wire grid mounted on the vacuum window.


\begin{figure}[ht]
	\includegraphics{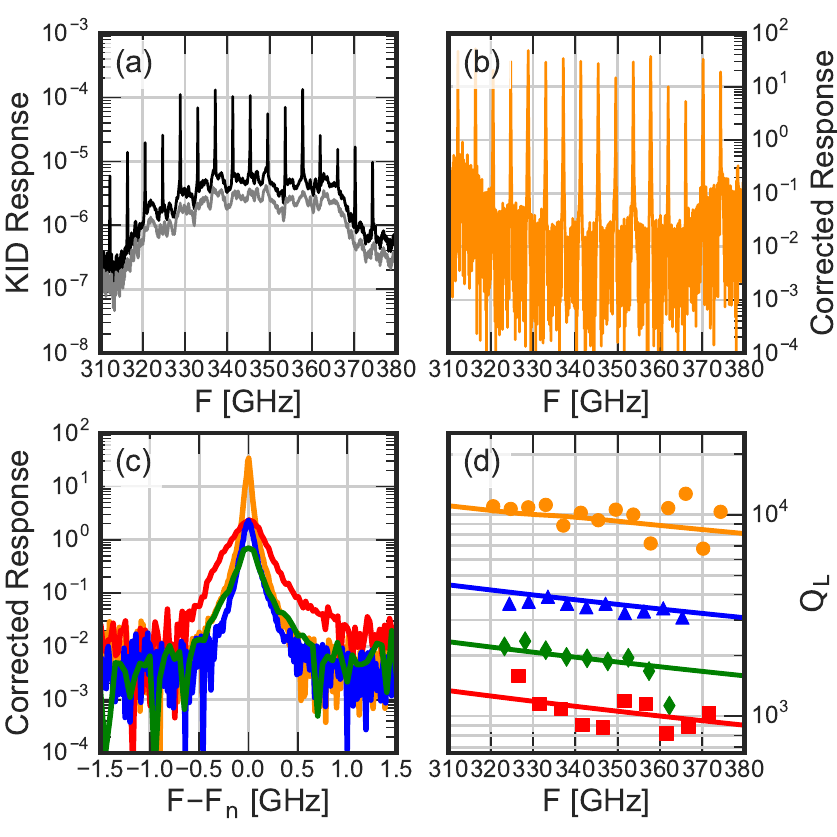}
	\caption{(a) Measured spectrum for chip I of the Fabry-P{\'e}rot coupled MKID (\textit{black}) and the blind MKID (\textit{grey}).
		(b) Corrected spectrum of the FP coupled MKID of chip I.
		(c) Example peaks of the corrected spectra of each chip (I: \textit{orange}; II: \textit{blue}; III: \textit{green}; IV: \textit{red}) 
		(d) Measured $Q_L$ for all chips and as symbols and simulated $Q_{L,sim}$ as lines, using the previous defined color scheme.}
	\label{fig:Response} 
\end{figure}

The FP transmission of the four chips is measured by sweeping the CW source from $310\ \mathrm{GHz}$ to $380\ \mathrm{GHz}$ in $10\ \mathrm{MHz}$ steps with an integration time of $1\ \mathrm{s}$ and detecting the resulting MKID response. An electrical on/off modulation of the CW source at $f_{mod}=11.97\ \mathrm{Hz}$ is employed to avoid 1/f noise. As the CW output power and beam shape are not well known, the absolute coupling strength to the MKID is not measured and the given responses are relative to the noise floor. However, the detector linearity in the measurement range was confirmed by measuring at various CW powers and retrieving identical results for the FP peak shapes. 

The resulting response $S_D$ of the FP-coupled MKID, shown in Fig. \ref{fig:Response}a) exemplary for chip I, clearly shows the expected regular spaced peaks of the FP resonator combined with a strongly frequency dependent baseline. The blind MKIDs spectrum $S_B$ shows the same baseline but with a frequency independent offset $O$ compared to $S_D$. As the same baseline is present in both detectors, we attribute it to CW power directly coupling to the MKIDs. Its frequency dependence is given by inherent fluctuations of the CW source combined with the bandpass filter transmission, both of which are also present in the Fabry-P{\'e}rot transmission, while the constant offset is due to the difference in MKID responsivity. We retrieve the corrected FP transmission $S_{FP}$ shown in Fig. \ref{fig:Response}b) using 
\begin{equation}
S_{FP} = S_{D}/S_{B} - O
\end{equation}
where $O$ is determined in the regions between FP peaks where $S_D$ is dominated by the direct coupling.

A comparison between the FP peaks of the 4 chips (see Fig. \ref{fig:Response}c)) shows sharper and higher peaks for narrower CPWs. This already indicates lower losses for the narrow CPWs, as the experiments were designed for the $Q_i$-limited regime ($Q_L \approx Q_i$). The peak height difference for chip IV is due to the use of a different aperture, which only affects the direct CW coupling and not the resonance Q factor.

In order to obtain $Q_L$ the individual peaks are fitted with a Lorentzian function
\begin{equation}
	L_n(F)=I\frac{Q_{L,n}^2}{Q_{L,n}^2+4\left(\frac{F-F_n}{F_n}\right)^2}+O_L
\end{equation}
with peak height $I$ and offset $O_L$, and the fit results plugged into Eqn.\ref{eq: Ql}. The fitted $Q_L$ is shown in Fig. \ref{fig:Response}d) as dots and compared to simulations shown by lines. 

The simulations are carried out in Sonnet (see the supplementary material) and are based on the measured CPW geometry and NbTiN properties as discussed previously. An excellent agreement with the measured data is found by including the coupling strength $S_{2'1'}$ and radiation loss of the CPW in these simulations. The observed frequency dependence of $Q_L$ is due to both the changing coupling strength and line loss $\alpha\propto f^3$, while the oscillation in measured $Q_L$ can be explained qualitatively by a standing wave before the first FP coupler with a resonance length $L_{sw} > L_{FP}$.

\begin{figure}
	\includegraphics{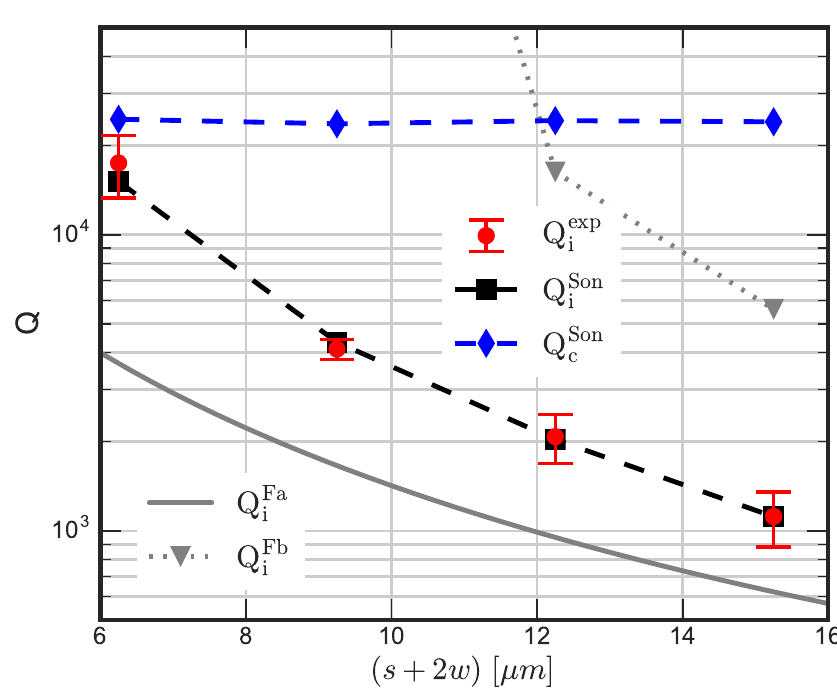}
	\caption{Measured $Q_i^{exp}$ compared with sonnet simulations for internal loss $Q_i^{Son}$ and coupling strength $Q_c^{chip}$, as well as analytical solutions for a PEC CPW $Q_i^{Fa}$ and superconducting CPW $Q_i^{Fb}$. }
	
	\label{fig:Qname}
\end{figure} 

In order to extract the internal loss from the measured $Q_L$, the $Q_c$ must be known. While it is in principle possible to measure $Q_c$ directly using the analysis in Fig. \ref{fig:FP}e), this requires a dynamic range $>50\ $dB or an absolute calibration of the $S_{21}$ at the resonance peaks, both of which are not possible in our experimental system. Therefore, we use the Sonnet simulations of the coupler to obtain $Q_c$. 

We then average over all peaks in the frequency range to retrieve $Q_i^{exp}$ at $350\ \mathrm{GHz}$, shown in Fig. \ref{fig:Qname},  which is in excellent agreement with the Sonnet simulations of the CPW radiation loss. It is significantly higher than the analytical solution $Q_i^{Fa}$ for the case of a PEC using Eqn.\ref{eq:frankel}, with the difference increasing for narrower lines up to a factor of 4. However, it is also significantly lower than the naive approach of substituting the superconducting $\epsilon_{eff}$ of Eqn. \ref{eq:eeff_sc} into Eqn.\ref{eq:frankel} resulting in $Q_i^{Fb}$. All four CPW geometries are within the validity range for Eqn.\ref{eq:frankel}, but as the derivation of Eqn.\ref{eq:frankel} is based on a planar PEC geometry, and does not take a superconductor into account, it is not surprising that both of these approaches fail. For $Q_i^{Fa}$ the phase velocity change due to the kinetic inductance is completely neglected, while the naive inclusion of $L_k$ in $\epsilon_{eff}$ for $Q_i^{Fb}$, while correct for phase velocity considerations, does not take into account the actual field distribution in the dielectric. 
The $Q_i$ of the NbTiN microwave resonator located on each of the chips is measured to be $\approx2\times10^6$, which is consistent with previous experiments \cite{Barends2010} and indicates no issues with film quality.

In addition to the quantitative disagreement between the experiment and the analytic solution, we find a non-zero loss for chip I where we expect no radiation loss according to the shockwave model. As $\epsilon_{eff}>\epsilon_{r}$ is a fundamental argument against radiation loss due to a shockwave, a different mechanism must be considered.
Radiation loss due to the strong fields at the open ended couplers were found to have a negligible contribution in sonnet simulations with $Q\approx10^5$ (see supplementary material). Dielectric losses due to the crystalline sapphire substrate are expected to be negligible and can be excluded due to the high $Q_i$ of the microwave resonator. Ohmic losses due to disorder effects in the NbTiN film are expected to be much smaller than observed and are not compatible with the measured width dependence. Additionally, none of these losses are included in the simulation for $Q_i^{son}$, where we find a quantitative agreement with the measurements. Due to this excellent agreement, we speculate that we are limited by a different loss mechanism, most likely due to the fundamentally unconfined nature of the CPW mode.

In conclusion, we have designed, fabricated and measured superconducting on-chip CPW Fabry-P{\'e}rot resonators with high kinetic inductance NbTiN ($L_s = 1.03/ $pH) and multiple line dimensions at frequencies from $320$ to $380\ \mathrm{GHz}$. We find a line width dependence for the internal loss $Q_i$, with values of $(1.1\pm0.2)\times10^3$ for a total line width of $15.25\ \mathrm{\mu m}$ to $(1.7\pm0.4)\times10^4$ for $6.25\ \mathrm{\mu m}$, corresponding to $\alpha=0.007\ \mathrm{dB/mm}$ and $\alpha=0.09\ \mathrm{dB/mm}$ respectively.The measured loss is in quantitative agreement with simulations of the radiation loss using Sonnet. However, the analytical solution by Frankel et al. \cite{Frankel92} is not valid in the regime of high-kinetic inductance superconductors, underestimating the CPW loss when $\epsilon_{eff} \approx \epsilon_r$.

Furthermore, we show that the on-chip Fabry-P{\'e}rot resonator provides a sensitive and highly flexible method for high-$Q_i$ transmission line loss measurements at sub-mm wavelengths. Extensions to other transmission line types, such as microstrips, can be easily achieved by modifying the resonator line and couplers, while the antenna can be exchanged to fit the required frequency range. Further optimization in the quasi-optical path and chip design are viable paths to improve the dynamic range and reduce effects from standing waves. For measurements of narrower lines where even lower losses are expected, a THz source with higher frequency resolution, such as multipliers, is required.

See supplementary material for the analytic characterization of a superconducting CPW and a comprehensive discussion of the Sonnet simulations for the Fabry-P{\'e}rot resonators.

The authors thank A. Neto for the helpful discussions.
This work is supported by the ERC COG 648135 MOSAIC. A. Endo, N.v. Marrewijk, and K. Karatsu were supported by the Netherlands Organization for Scientific Research NWO (Vidi grant No. 639.042.423)

The data that support the findings of this study are available from the corresponding author upon reasonable request.

\bibliography{Bibliography}

\providecommand{\noopsort}[1]{}\providecommand{\singleletter}[1]{#1}%
\begin{thebibliography}{19}%
\makeatletter
\providecommand \@ifxundefined [1]{%
 \@ifx{#1\undefined}
}%
\providecommand \@ifnum [1]{%
 \ifnum #1\expandafter \@firstoftwo
 \else \expandafter \@secondoftwo
 \fi
}%
\providecommand \@ifx [1]{%
 \ifx #1\expandafter \@firstoftwo
 \else \expandafter \@secondoftwo
 \fi
}%
\providecommand \natexlab [1]{#1}%
\providecommand \enquote  [1]{``#1''}%
\providecommand \bibnamefont  [1]{#1}%
\providecommand \bibfnamefont [1]{#1}%
\providecommand \citenamefont [1]{#1}%
\providecommand \href@noop [0]{\@secondoftwo}%
\providecommand \href [0]{\begingroup \@sanitize@url \@href}%
\providecommand \@href[1]{\@@startlink{#1}\@@href}%
\providecommand \@@href[1]{\endgroup#1\@@endlink}%
\providecommand \@sanitize@url [0]{\catcode `\\12\catcode `\$12\catcode
  `\&12\catcode `\#12\catcode `\^12\catcode `\_12\catcode `\%12\relax}%
\providecommand \@@startlink[1]{}%
\providecommand \@@endlink[0]{}%
\providecommand \url  [0]{\begingroup\@sanitize@url \@url }%
\providecommand \@url [1]{\endgroup\@href {#1}{\urlprefix }}%
\providecommand \urlprefix  [0]{URL }%
\providecommand \Eprint [0]{\href }%
\providecommand \doibase [0]{http://dx.doi.org/}%
\providecommand \selectlanguage [0]{\@gobble}%
\providecommand \bibinfo  [0]{\@secondoftwo}%
\providecommand \bibfield  [0]{\@secondoftwo}%
\providecommand \translation [1]{[#1]}%
\providecommand \BibitemOpen [0]{}%
\providecommand \bibitemStop [0]{}%
\providecommand \bibitemNoStop [0]{.\EOS\space}%
\providecommand \EOS [0]{\spacefactor3000\relax}%
\providecommand \BibitemShut  [1]{\csname bibitem#1\endcsname}%
\let\auto@bib@innerbib\@empty
\bibitem [{\citenamefont {{Endo}}\ \emph {et~al.}(2019)\citenamefont {{Endo}},
  \citenamefont {{Karatsu}}, \citenamefont {{Tamura}}, \citenamefont
  {{Oshima}}, \citenamefont {{Taniguchi}}, \citenamefont {{Takekoshi}},
  \citenamefont {{Asayama}}, \citenamefont {{Bakx}}, \citenamefont {{Bosma}},
  \citenamefont {{Bueno}}, \citenamefont {{Chin}}, \citenamefont {{Fujii}},
  \citenamefont {{Fujita}}, \citenamefont {{Huiting}}, \citenamefont
  {{Ikarashi}}, \citenamefont {{Ishida}}, \citenamefont {{Ishii}},
  \citenamefont {{Kawabe}}, \citenamefont {{Klapwijk}}, \citenamefont
  {{Kohno}}, \citenamefont {{Kouchi}}, \citenamefont {{Llombart}},
  \citenamefont {{Maekawa}}, \citenamefont {{Murugesan}}, \citenamefont
  {{Nakatsubo}}, \citenamefont {{Naruse}}, \citenamefont {{Ohtawara}},
  \citenamefont {{Pascual Laguna}}, \citenamefont {{Suzuki}}, \citenamefont
  {{Suzuki}}, \citenamefont {{Thoen}}, \citenamefont {{Tsukagoshi}},
  \citenamefont {{Ueda}}, \citenamefont {{de Visser}}, \citenamefont {{van der
  Werf}}, \citenamefont {{Yates}}, \citenamefont {{Yoshimura}}, \citenamefont
  {{Yurduseven}},\ and\ \citenamefont {{Baselmans}}}]{DeshimaNature}%
  \BibitemOpen
  \bibfield  {author} {\bibinfo {author} {\bibfnamefont {A.}~\bibnamefont
  {{Endo}}}, \bibinfo {author} {\bibfnamefont {K.}~\bibnamefont {{Karatsu}}},
  \bibinfo {author} {\bibfnamefont {Y.}~\bibnamefont {{Tamura}}}, \bibinfo
  {author} {\bibfnamefont {T.}~\bibnamefont {{Oshima}}}, \bibinfo {author}
  {\bibfnamefont {A.}~\bibnamefont {{Taniguchi}}}, \bibinfo {author}
  {\bibfnamefont {T.}~\bibnamefont {{Takekoshi}}}, \bibinfo {author}
  {\bibfnamefont {S.}~\bibnamefont {{Asayama}}}, \bibinfo {author}
  {\bibfnamefont {T.~J.~L.~C.}\ \bibnamefont {{Bakx}}}, \bibinfo {author}
  {\bibfnamefont {S.}~\bibnamefont {{Bosma}}}, \bibinfo {author} {\bibfnamefont
  {J.}~\bibnamefont {{Bueno}}}, \bibinfo {author} {\bibfnamefont {K.~W.}\
  \bibnamefont {{Chin}}}, \bibinfo {author} {\bibfnamefont {Y.}~\bibnamefont
  {{Fujii}}}, \bibinfo {author} {\bibfnamefont {K.}~\bibnamefont {{Fujita}}},
  \bibinfo {author} {\bibfnamefont {R.}~\bibnamefont {{Huiting}}}, \bibinfo
  {author} {\bibfnamefont {S.}~\bibnamefont {{Ikarashi}}}, \bibinfo {author}
  {\bibfnamefont {T.}~\bibnamefont {{Ishida}}}, \bibinfo {author}
  {\bibfnamefont {S.}~\bibnamefont {{Ishii}}}, \bibinfo {author} {\bibfnamefont
  {R.}~\bibnamefont {{Kawabe}}}, \bibinfo {author} {\bibfnamefont {T.~M.}\
  \bibnamefont {{Klapwijk}}}, \bibinfo {author} {\bibfnamefont
  {K.}~\bibnamefont {{Kohno}}}, \bibinfo {author} {\bibfnamefont
  {A.}~\bibnamefont {{Kouchi}}}, \bibinfo {author} {\bibfnamefont
  {N.}~\bibnamefont {{Llombart}}}, \bibinfo {author} {\bibfnamefont
  {J.}~\bibnamefont {{Maekawa}}}, \bibinfo {author} {\bibfnamefont
  {V.}~\bibnamefont {{Murugesan}}}, \bibinfo {author} {\bibfnamefont
  {S.}~\bibnamefont {{Nakatsubo}}}, \bibinfo {author} {\bibfnamefont
  {M.}~\bibnamefont {{Naruse}}}, \bibinfo {author} {\bibfnamefont
  {K.}~\bibnamefont {{Ohtawara}}}, \bibinfo {author} {\bibfnamefont
  {A.}~\bibnamefont {{Pascual Laguna}}}, \bibinfo {author} {\bibfnamefont
  {J.}~\bibnamefont {{Suzuki}}}, \bibinfo {author} {\bibfnamefont
  {K.}~\bibnamefont {{Suzuki}}}, \bibinfo {author} {\bibfnamefont {D.~J.}\
  \bibnamefont {{Thoen}}}, \bibinfo {author} {\bibfnamefont {T.}~\bibnamefont
  {{Tsukagoshi}}}, \bibinfo {author} {\bibfnamefont {T.}~\bibnamefont
  {{Ueda}}}, \bibinfo {author} {\bibfnamefont {P.~J.}\ \bibnamefont {{de
  Visser}}}, \bibinfo {author} {\bibfnamefont {P.~P.}\ \bibnamefont {{van der
  Werf}}}, \bibinfo {author} {\bibfnamefont {S.~J.~C.}\ \bibnamefont
  {{Yates}}}, \bibinfo {author} {\bibfnamefont {Y.}~\bibnamefont
  {{Yoshimura}}}, \bibinfo {author} {\bibfnamefont {O.}~\bibnamefont
  {{Yurduseven}}}, \ and\ \bibinfo {author} {\bibfnamefont {J.~J.~A.}\
  \bibnamefont {{Baselmans}}},\ }\href {\doibase 10.1038/s41550-019-0850-8}
  {\bibfield  {journal} {\bibinfo  {journal} {Nature Astronomy}\ }\textbf
  {\bibinfo {volume} {3}},\ \bibinfo {pages} {989} (\bibinfo {year} {2019})},\
  \Eprint {http://arxiv.org/abs/1906.10216} {arXiv:1906.10216 [astro-ph.IM]}
  \BibitemShut {NoStop}%
\bibitem [{\citenamefont {Cataldo}\ \emph {et~al.}(2018)\citenamefont
  {Cataldo}, \citenamefont {Barrentine}, \citenamefont {Bulcha}, \citenamefont
  {Ehsan}, \citenamefont {Hess}, \citenamefont {Noroozian}, \citenamefont
  {Stevenson}, \citenamefont {U-Yen}, \citenamefont {Wollack},\ and\
  \citenamefont {Moseley}}]{Cataldo2018}%
  \BibitemOpen
  \bibfield  {author} {\bibinfo {author} {\bibfnamefont {G.}~\bibnamefont
  {Cataldo}}, \bibinfo {author} {\bibfnamefont {E.~M.}\ \bibnamefont
  {Barrentine}}, \bibinfo {author} {\bibfnamefont {B.~T.}\ \bibnamefont
  {Bulcha}}, \bibinfo {author} {\bibfnamefont {N.}~\bibnamefont {Ehsan}},
  \bibinfo {author} {\bibfnamefont {L.~A.}\ \bibnamefont {Hess}}, \bibinfo
  {author} {\bibfnamefont {O.}~\bibnamefont {Noroozian}}, \bibinfo {author}
  {\bibfnamefont {T.~R.}\ \bibnamefont {Stevenson}}, \bibinfo {author}
  {\bibfnamefont {K.}~\bibnamefont {U-Yen}}, \bibinfo {author} {\bibfnamefont
  {E.~J.}\ \bibnamefont {Wollack}}, \ and\ \bibinfo {author} {\bibfnamefont
  {S.~H.}\ \bibnamefont {Moseley}},\ }\href {\doibase
  10.1007/s10909-018-1902-7} {\bibfield  {journal} {\bibinfo  {journal}
  {Journal of Low Temperature Physics}\ }\textbf {\bibinfo {volume} {193}},\
  \bibinfo {pages} {923} (\bibinfo {year} {2018})}\BibitemShut {NoStop}%
\bibitem [{\citenamefont {Shirokoff}\ \emph {et~al.}(2012)\citenamefont
  {Shirokoff}, \citenamefont {Barry}, \citenamefont {Bradford}, \citenamefont
  {Chattopadhyay}, \citenamefont {Day}, \citenamefont {Doyle}, \citenamefont
  {Hailey-Dunsheath}, \citenamefont {Hollister}, \citenamefont {Kovács},
  \citenamefont {McKenney}, \citenamefont {Leduc}, \citenamefont {Llombart},
  \citenamefont {Marrone}, \citenamefont {Mauskopf}, \citenamefont {O'Brient},
  \citenamefont {Padin}, \citenamefont {Reck}, \citenamefont {Swenson},\ and\
  \citenamefont {Zmuidzinas}}]{Superspec}%
  \BibitemOpen
  \bibfield  {author} {\bibinfo {author} {\bibfnamefont {E.}~\bibnamefont
  {Shirokoff}}, \bibinfo {author} {\bibfnamefont {P.~S.}\ \bibnamefont
  {Barry}}, \bibinfo {author} {\bibfnamefont {C.~M.}\ \bibnamefont {Bradford}},
  \bibinfo {author} {\bibfnamefont {G.}~\bibnamefont {Chattopadhyay}}, \bibinfo
  {author} {\bibfnamefont {P.}~\bibnamefont {Day}}, \bibinfo {author}
  {\bibfnamefont {S.}~\bibnamefont {Doyle}}, \bibinfo {author} {\bibfnamefont
  {S.}~\bibnamefont {Hailey-Dunsheath}}, \bibinfo {author} {\bibfnamefont
  {M.~I.}\ \bibnamefont {Hollister}}, \bibinfo {author} {\bibfnamefont
  {A.}~\bibnamefont {Kovács}}, \bibinfo {author} {\bibfnamefont
  {C.}~\bibnamefont {McKenney}}, \bibinfo {author} {\bibfnamefont {H.~G.}\
  \bibnamefont {Leduc}}, \bibinfo {author} {\bibfnamefont {N.}~\bibnamefont
  {Llombart}}, \bibinfo {author} {\bibfnamefont {D.~P.}\ \bibnamefont
  {Marrone}}, \bibinfo {author} {\bibfnamefont {P.}~\bibnamefont {Mauskopf}},
  \bibinfo {author} {\bibfnamefont {R.}~\bibnamefont {O'Brient}}, \bibinfo
  {author} {\bibfnamefont {S.}~\bibnamefont {Padin}}, \bibinfo {author}
  {\bibfnamefont {T.}~\bibnamefont {Reck}}, \bibinfo {author} {\bibfnamefont
  {L.~J.}\ \bibnamefont {Swenson}}, \ and\ \bibinfo {author} {\bibfnamefont
  {J.}~\bibnamefont {Zmuidzinas}},\ }in\ \href {\doibase 10.1117/12.927070}
  {\emph {\bibinfo {booktitle} {Millimeter, Submillimeter, and Far-Infrared
  Detectors and Instrumentation for Astronomy VI}}},\ Vol.\ \bibinfo {volume}
  {8452},\ \bibinfo {editor} {edited by\ \bibinfo {editor} {\bibfnamefont
  {W.~S.}\ \bibnamefont {Holland}}},\ \bibinfo {organization} {International
  Society for Optics and Photonics}\ (\bibinfo  {publisher} {SPIE},\ \bibinfo
  {year} {2012})\ pp.\ \bibinfo {pages} {209 -- 219}\BibitemShut {NoStop}%
\bibitem [{\citenamefont {Ade}\ \emph {et~al.}(2015)\citenamefont {Ade},
  \citenamefont {Aikin}, \citenamefont {Amiri}, \citenamefont {Barkats},
  \citenamefont {Benton}, \citenamefont {Bischoff}, \citenamefont {Bock},
  \citenamefont {Bonetti}, \citenamefont {Brevik}, \citenamefont {Buder},
  \citenamefont {Bullock}, \citenamefont {Chattopadhyay}, \citenamefont
  {Davis}, \citenamefont {Day}, \citenamefont {Dowell}, \citenamefont {Duband},
  \citenamefont {Filippini}, \citenamefont {Fliescher}, \citenamefont
  {Golwala}, \citenamefont {Halpern}, \citenamefont {Hasselfield},
  \citenamefont {Hildebrandt}, \citenamefont {Hilton}, \citenamefont {Hristov},
  \citenamefont {Hui}, \citenamefont {Irwin}, \citenamefont {Jones},
  \citenamefont {Karkare}, \citenamefont {Kaufman}, \citenamefont {Keating},
  \citenamefont {Kefeli}, \citenamefont {Kernasovskiy}, \citenamefont {Kovac},
  \citenamefont {Kuo}, \citenamefont {LeDuc}, \citenamefont {Leitch},
  \citenamefont {Llombart}, \citenamefont {Lueker}, \citenamefont {Mason},
  \citenamefont {Megerian}, \citenamefont {Moncelsi}, \citenamefont
  {Netterfield}, \citenamefont {Nguyen}, \citenamefont {O'Brient},
  \citenamefont {IV}, \citenamefont {Orlando}, \citenamefont {Pryke},
  \citenamefont {Rahlin}, \citenamefont {Reintsema}, \citenamefont {Richter},
  \citenamefont {Runyan}, \citenamefont {Schwarz}, \citenamefont {Sheehy},
  \citenamefont {Staniszewski}, \citenamefont {Sudiwala}, \citenamefont
  {Teply}, \citenamefont {Tolan}, \citenamefont {Trangsrud}, \citenamefont
  {Tucker}, \citenamefont {Turner}, \citenamefont {Vieregg}, \citenamefont
  {Weber}, \citenamefont {Wiebe}, \citenamefont {Wilson}, \citenamefont {Wong},
  \citenamefont {Yoon},\ and\ \citenamefont {and}}]{Ade_2015}%
  \BibitemOpen
  \bibfield  {author} {\bibinfo {author} {\bibfnamefont {P.~A.~R.}\
  \bibnamefont {Ade}}, \bibinfo {author} {\bibfnamefont {R.~W.}\ \bibnamefont
  {Aikin}}, \bibinfo {author} {\bibfnamefont {M.}~\bibnamefont {Amiri}},
  \bibinfo {author} {\bibfnamefont {D.}~\bibnamefont {Barkats}}, \bibinfo
  {author} {\bibfnamefont {S.~J.}\ \bibnamefont {Benton}}, \bibinfo {author}
  {\bibfnamefont {C.~A.}\ \bibnamefont {Bischoff}}, \bibinfo {author}
  {\bibfnamefont {J.~J.}\ \bibnamefont {Bock}}, \bibinfo {author}
  {\bibfnamefont {J.~A.}\ \bibnamefont {Bonetti}}, \bibinfo {author}
  {\bibfnamefont {J.~A.}\ \bibnamefont {Brevik}}, \bibinfo {author}
  {\bibfnamefont {I.}~\bibnamefont {Buder}}, \bibinfo {author} {\bibfnamefont
  {E.}~\bibnamefont {Bullock}}, \bibinfo {author} {\bibfnamefont
  {G.}~\bibnamefont {Chattopadhyay}}, \bibinfo {author} {\bibfnamefont
  {G.}~\bibnamefont {Davis}}, \bibinfo {author} {\bibfnamefont {P.~K.}\
  \bibnamefont {Day}}, \bibinfo {author} {\bibfnamefont {C.~D.}\ \bibnamefont
  {Dowell}}, \bibinfo {author} {\bibfnamefont {L.}~\bibnamefont {Duband}},
  \bibinfo {author} {\bibfnamefont {J.~P.}\ \bibnamefont {Filippini}}, \bibinfo
  {author} {\bibfnamefont {S.}~\bibnamefont {Fliescher}}, \bibinfo {author}
  {\bibfnamefont {S.~R.}\ \bibnamefont {Golwala}}, \bibinfo {author}
  {\bibfnamefont {M.}~\bibnamefont {Halpern}}, \bibinfo {author} {\bibfnamefont
  {M.}~\bibnamefont {Hasselfield}}, \bibinfo {author} {\bibfnamefont {S.~R.}\
  \bibnamefont {Hildebrandt}}, \bibinfo {author} {\bibfnamefont {G.~C.}\
  \bibnamefont {Hilton}}, \bibinfo {author} {\bibfnamefont {V.}~\bibnamefont
  {Hristov}}, \bibinfo {author} {\bibfnamefont {H.}~\bibnamefont {Hui}},
  \bibinfo {author} {\bibfnamefont {K.~D.}\ \bibnamefont {Irwin}}, \bibinfo
  {author} {\bibfnamefont {W.~C.}\ \bibnamefont {Jones}}, \bibinfo {author}
  {\bibfnamefont {K.~S.}\ \bibnamefont {Karkare}}, \bibinfo {author}
  {\bibfnamefont {J.~P.}\ \bibnamefont {Kaufman}}, \bibinfo {author}
  {\bibfnamefont {B.~G.}\ \bibnamefont {Keating}}, \bibinfo {author}
  {\bibfnamefont {S.}~\bibnamefont {Kefeli}}, \bibinfo {author} {\bibfnamefont
  {S.~A.}\ \bibnamefont {Kernasovskiy}}, \bibinfo {author} {\bibfnamefont
  {J.~M.}\ \bibnamefont {Kovac}}, \bibinfo {author} {\bibfnamefont {C.~L.}\
  \bibnamefont {Kuo}}, \bibinfo {author} {\bibfnamefont {H.~G.}\ \bibnamefont
  {LeDuc}}, \bibinfo {author} {\bibfnamefont {E.~M.}\ \bibnamefont {Leitch}},
  \bibinfo {author} {\bibfnamefont {N.}~\bibnamefont {Llombart}}, \bibinfo
  {author} {\bibfnamefont {M.}~\bibnamefont {Lueker}}, \bibinfo {author}
  {\bibfnamefont {P.}~\bibnamefont {Mason}}, \bibinfo {author} {\bibfnamefont
  {K.}~\bibnamefont {Megerian}}, \bibinfo {author} {\bibfnamefont
  {L.}~\bibnamefont {Moncelsi}}, \bibinfo {author} {\bibfnamefont {C.~B.}\
  \bibnamefont {Netterfield}}, \bibinfo {author} {\bibfnamefont {H.~T.}\
  \bibnamefont {Nguyen}}, \bibinfo {author} {\bibfnamefont {R.}~\bibnamefont
  {O'Brient}}, \bibinfo {author} {\bibfnamefont {R.~W.~O.}\ \bibnamefont {IV}},
  \bibinfo {author} {\bibfnamefont {A.}~\bibnamefont {Orlando}}, \bibinfo
  {author} {\bibfnamefont {C.}~\bibnamefont {Pryke}}, \bibinfo {author}
  {\bibfnamefont {A.~S.}\ \bibnamefont {Rahlin}}, \bibinfo {author}
  {\bibfnamefont {C.~D.}\ \bibnamefont {Reintsema}}, \bibinfo {author}
  {\bibfnamefont {S.}~\bibnamefont {Richter}}, \bibinfo {author} {\bibfnamefont
  {M.~C.}\ \bibnamefont {Runyan}}, \bibinfo {author} {\bibfnamefont
  {R.}~\bibnamefont {Schwarz}}, \bibinfo {author} {\bibfnamefont {C.~D.}\
  \bibnamefont {Sheehy}}, \bibinfo {author} {\bibfnamefont {Z.~K.}\
  \bibnamefont {Staniszewski}}, \bibinfo {author} {\bibfnamefont {R.~V.}\
  \bibnamefont {Sudiwala}}, \bibinfo {author} {\bibfnamefont {G.~P.}\
  \bibnamefont {Teply}}, \bibinfo {author} {\bibfnamefont {J.~E.}\ \bibnamefont
  {Tolan}}, \bibinfo {author} {\bibfnamefont {A.}~\bibnamefont {Trangsrud}},
  \bibinfo {author} {\bibfnamefont {R.~S.}\ \bibnamefont {Tucker}}, \bibinfo
  {author} {\bibfnamefont {A.~D.}\ \bibnamefont {Turner}}, \bibinfo {author}
  {\bibfnamefont {A.~G.}\ \bibnamefont {Vieregg}}, \bibinfo {author}
  {\bibfnamefont {A.}~\bibnamefont {Weber}}, \bibinfo {author} {\bibfnamefont
  {D.~V.}\ \bibnamefont {Wiebe}}, \bibinfo {author} {\bibfnamefont
  {P.}~\bibnamefont {Wilson}}, \bibinfo {author} {\bibfnamefont {C.~L.}\
  \bibnamefont {Wong}}, \bibinfo {author} {\bibfnamefont {K.~W.}\ \bibnamefont
  {Yoon}}, \ and\ \bibinfo {author} {\bibfnamefont {J.~Z.}\ \bibnamefont
  {and}},\ }\href {\doibase 10.1088/0004-637x/812/2/176} {\bibfield  {journal}
  {\bibinfo  {journal} {The Astrophysical Journal}\ }\textbf {\bibinfo {volume}
  {812}},\ \bibinfo {pages} {176} (\bibinfo {year} {2015})}\BibitemShut
  {NoStop}%
\bibitem [{\citenamefont {Ho~Eom}\ \emph {et~al.}(2012)\citenamefont {Ho~Eom},
  \citenamefont {Day}, \citenamefont {LeDuc},\ and\ \citenamefont
  {Zmuidzinas}}]{ho_eom_wideband_2012}%
  \BibitemOpen
  \bibfield  {author} {\bibinfo {author} {\bibfnamefont {B.}~\bibnamefont
  {Ho~Eom}}, \bibinfo {author} {\bibfnamefont {P.~K.}\ \bibnamefont {Day}},
  \bibinfo {author} {\bibfnamefont {H.~G.}\ \bibnamefont {LeDuc}}, \ and\
  \bibinfo {author} {\bibfnamefont {J.}~\bibnamefont {Zmuidzinas}},\ }\href
  {\doibase 10.1038/nphys2356} {\bibfield  {journal} {\bibinfo  {journal}
  {Nature Physics}\ }\textbf {\bibinfo {volume} {8}},\ \bibinfo {pages} {623}
  (\bibinfo {year} {2012})}\BibitemShut {NoStop}%
\bibitem [{\citenamefont {Gao}\ \emph {et~al.}(2009)\citenamefont {Gao},
  \citenamefont {Vayonakis}, \citenamefont {Noroozian}, \citenamefont
  {Zmuidzinas}, \citenamefont {Day},\ and\ \citenamefont {Leduc}}]{Gao2009}%
  \BibitemOpen
  \bibfield  {author} {\bibinfo {author} {\bibfnamefont {J.}~\bibnamefont
  {Gao}}, \bibinfo {author} {\bibfnamefont {A.}~\bibnamefont {Vayonakis}},
  \bibinfo {author} {\bibfnamefont {O.}~\bibnamefont {Noroozian}}, \bibinfo
  {author} {\bibfnamefont {J.}~\bibnamefont {Zmuidzinas}}, \bibinfo {author}
  {\bibfnamefont {P.}~\bibnamefont {Day}}, \ and\ \bibinfo {author}
  {\bibfnamefont {H.}~\bibnamefont {Leduc}},\ }\href {\doibase
  10.1063/1.3292306} {\ \textbf {\bibinfo {volume} {1185}} (\bibinfo {year}
  {2009}),\ 10.1063/1.3292306}\BibitemShut {NoStop}%
\bibitem [{\citenamefont {{Frankel}}\ \emph {et~al.}(1991)\citenamefont
  {{Frankel}}, \citenamefont {{Gupta}}, \citenamefont {{Valdmanis}},\ and\
  \citenamefont {{Mourou}}}]{Frankel92}%
  \BibitemOpen
  \bibfield  {author} {\bibinfo {author} {\bibfnamefont {M.~Y.}\ \bibnamefont
  {{Frankel}}}, \bibinfo {author} {\bibfnamefont {S.}~\bibnamefont {{Gupta}}},
  \bibinfo {author} {\bibfnamefont {J.~A.}\ \bibnamefont {{Valdmanis}}}, \ and\
  \bibinfo {author} {\bibfnamefont {G.~A.}\ \bibnamefont {{Mourou}}},\ }\href
  {\doibase 10.1109/22.81658} {\bibfield  {journal} {\bibinfo  {journal} {IEEE
  Transactions on Microwave Theory and Techniques}\ }\textbf {\bibinfo {volume}
  {39}},\ \bibinfo {pages} {910} (\bibinfo {year} {1991})}\BibitemShut
  {NoStop}%
\bibitem [{\citenamefont {Göppl}\ \emph {et~al.}(2008)\citenamefont {Göppl},
  \citenamefont {Fragner}, \citenamefont {Baur}, \citenamefont {Bianchetti},
  \citenamefont {Filipp}, \citenamefont {Fink}, \citenamefont {Leek},
  \citenamefont {Puebla}, \citenamefont {Steffen},\ and\ \citenamefont
  {Wallraff}}]{Goeppl08}%
  \BibitemOpen
  \bibfield  {author} {\bibinfo {author} {\bibfnamefont {M.}~\bibnamefont
  {Göppl}}, \bibinfo {author} {\bibfnamefont {A.}~\bibnamefont {Fragner}},
  \bibinfo {author} {\bibfnamefont {M.}~\bibnamefont {Baur}}, \bibinfo {author}
  {\bibfnamefont {R.}~\bibnamefont {Bianchetti}}, \bibinfo {author}
  {\bibfnamefont {S.}~\bibnamefont {Filipp}}, \bibinfo {author} {\bibfnamefont
  {J.~M.}\ \bibnamefont {Fink}}, \bibinfo {author} {\bibfnamefont {P.~J.}\
  \bibnamefont {Leek}}, \bibinfo {author} {\bibfnamefont {G.}~\bibnamefont
  {Puebla}}, \bibinfo {author} {\bibfnamefont {L.}~\bibnamefont {Steffen}}, \
  and\ \bibinfo {author} {\bibfnamefont {A.}~\bibnamefont {Wallraff}},\ }\href
  {\doibase 10.1063/1.3010859} {\bibfield  {journal} {\bibinfo  {journal}
  {Journal of Applied Physics}\ }\textbf {\bibinfo {volume} {104}},\ \bibinfo
  {pages} {113904} (\bibinfo {year} {2008})},\ \Eprint
  {http://arxiv.org/abs/https://doi.org/10.1063/1.3010859}
  {https://doi.org/10.1063/1.3010859} \BibitemShut {NoStop}%
\bibitem [{\citenamefont {Manual}(2008)}]{manual2008sonnet}%
  \BibitemOpen
  \bibfield  {author} {\bibinfo {author} {\bibfnamefont {E.~U.}\ \bibnamefont
  {Manual}},\ }\href@noop {} {\bibfield  {journal} {\bibinfo  {journal} {Inc.,
  Liverpool, NY}\ } (\bibinfo {year} {2008})}\BibitemShut {NoStop}%
\bibitem [{\citenamefont {{Thoen}}\ \emph {et~al.}(2017)\citenamefont
  {{Thoen}}, \citenamefont {{Bos}}, \citenamefont {{Haalebos}}, \citenamefont
  {{Klapwijk}}, \citenamefont {{Baselmans}},\ and\ \citenamefont
  {{Endo}}}]{Thoen_NbTiN}%
  \BibitemOpen
  \bibfield  {author} {\bibinfo {author} {\bibfnamefont {D.~J.}\ \bibnamefont
  {{Thoen}}}, \bibinfo {author} {\bibfnamefont {B.~G.~C.}\ \bibnamefont
  {{Bos}}}, \bibinfo {author} {\bibfnamefont {E.~A.~F.}\ \bibnamefont
  {{Haalebos}}}, \bibinfo {author} {\bibfnamefont {T.~M.}\ \bibnamefont
  {{Klapwijk}}}, \bibinfo {author} {\bibfnamefont {J.~J.~A.}\ \bibnamefont
  {{Baselmans}}}, \ and\ \bibinfo {author} {\bibfnamefont {A.}~\bibnamefont
  {{Endo}}},\ }\href {\doibase 10.1109/TASC.2016.2631948} {\bibfield  {journal}
  {\bibinfo  {journal} {IEEE Transactions on Applied Superconductivity}\
  }\textbf {\bibinfo {volume} {27}},\ \bibinfo {pages} {1} (\bibinfo {year}
  {2017})}\BibitemShut {NoStop}%
\bibitem [{\citenamefont {Endo}\ \emph {et~al.}(2019)\citenamefont {Endo},
  \citenamefont {Karatsu}, \citenamefont {Laguna}, \citenamefont {Mirzaei},
  \citenamefont {Huiting}, \citenamefont {Thoen}, \citenamefont {Murugesan},
  \citenamefont {Yates}, \citenamefont {Bueno}, \citenamefont {Marrewijk},
  \citenamefont {Bosma}, \citenamefont {Yurduseven}, \citenamefont {Llombart},
  \citenamefont {Suzuki}, \citenamefont {Naruse}, \citenamefont {de~Visser},
  \citenamefont {van~der Werf}, \citenamefont {Klapwijk},\ and\ \citenamefont
  {Baselmans}}]{EndoJatis}%
  \BibitemOpen
  \bibfield  {author} {\bibinfo {author} {\bibfnamefont {A.}~\bibnamefont
  {Endo}}, \bibinfo {author} {\bibfnamefont {K.}~\bibnamefont {Karatsu}},
  \bibinfo {author} {\bibfnamefont {A.~P.}\ \bibnamefont {Laguna}}, \bibinfo
  {author} {\bibfnamefont {B.}~\bibnamefont {Mirzaei}}, \bibinfo {author}
  {\bibfnamefont {R.}~\bibnamefont {Huiting}}, \bibinfo {author} {\bibfnamefont
  {D.}~\bibnamefont {Thoen}}, \bibinfo {author} {\bibfnamefont
  {V.}~\bibnamefont {Murugesan}}, \bibinfo {author} {\bibfnamefont {S.~J.~C.}\
  \bibnamefont {Yates}}, \bibinfo {author} {\bibfnamefont {J.}~\bibnamefont
  {Bueno}}, \bibinfo {author} {\bibfnamefont {N.~V.}\ \bibnamefont
  {Marrewijk}}, \bibinfo {author} {\bibfnamefont {S.}~\bibnamefont {Bosma}},
  \bibinfo {author} {\bibfnamefont {O.}~\bibnamefont {Yurduseven}}, \bibinfo
  {author} {\bibfnamefont {N.}~\bibnamefont {Llombart}}, \bibinfo {author}
  {\bibfnamefont {J.}~\bibnamefont {Suzuki}}, \bibinfo {author} {\bibfnamefont
  {M.}~\bibnamefont {Naruse}}, \bibinfo {author} {\bibfnamefont {P.~J.}\
  \bibnamefont {de~Visser}}, \bibinfo {author} {\bibfnamefont {P.~P.}\
  \bibnamefont {van~der Werf}}, \bibinfo {author} {\bibfnamefont {T.~M.}\
  \bibnamefont {Klapwijk}}, \ and\ \bibinfo {author} {\bibfnamefont {J.~J.~A.}\
  \bibnamefont {Baselmans}},\ }\href {\doibase 10.1117/1.JATIS.5.3.035004}
  {\bibfield  {journal} {\bibinfo  {journal} {Journal of Astronomical
  Telescopes, Instruments, and Systems}\ }\textbf {\bibinfo {volume} {5}},\
  \bibinfo {pages} {1 } (\bibinfo {year} {2019})}\BibitemShut {NoStop}%
\bibitem [{\citenamefont {Mattis}\ and\ \citenamefont
  {Bardeen}(1958)}]{MattisBardeen}%
  \BibitemOpen
  \bibfield  {author} {\bibinfo {author} {\bibfnamefont {D.~C.}\ \bibnamefont
  {Mattis}}\ and\ \bibinfo {author} {\bibfnamefont {J.}~\bibnamefont
  {Bardeen}},\ }\href {\doibase 10.1103/PhysRev.111.412} {\bibfield  {journal}
  {\bibinfo  {journal} {Phys. Rev.}\ }\textbf {\bibinfo {volume} {111}},\
  \bibinfo {pages} {412} (\bibinfo {year} {1958})}\BibitemShut {NoStop}%
\bibitem [{\citenamefont {Driessen}\ \emph {et~al.}(2012)\citenamefont
  {Driessen}, \citenamefont {Coumou}, \citenamefont {Tromp}, \citenamefont
  {de~Visser},\ and\ \citenamefont {Klapwijk}}]{Driessen2012_PRL}%
  \BibitemOpen
  \bibfield  {author} {\bibinfo {author} {\bibfnamefont {E.~F.~C.}\
  \bibnamefont {Driessen}}, \bibinfo {author} {\bibfnamefont {P.~C. J.~J.}\
  \bibnamefont {Coumou}}, \bibinfo {author} {\bibfnamefont {R.~R.}\
  \bibnamefont {Tromp}}, \bibinfo {author} {\bibfnamefont {P.~J.}\ \bibnamefont
  {de~Visser}}, \ and\ \bibinfo {author} {\bibfnamefont {T.~M.}\ \bibnamefont
  {Klapwijk}},\ }\href {\doibase 10.1103/PhysRevLett.109.107003} {\bibfield
  {journal} {\bibinfo  {journal} {Phys. Rev. Lett.}\ }\textbf {\bibinfo
  {volume} {109}},\ \bibinfo {pages} {107003} (\bibinfo {year}
  {2012})}\BibitemShut {NoStop}%
\bibitem [{\citenamefont {Janssen}\ \emph {et~al.}(2013)\citenamefont
  {Janssen}, \citenamefont {Baselmans}, \citenamefont {Endo}, \citenamefont
  {Ferrari}, \citenamefont {Yates}, \citenamefont {Baryshev},\ and\
  \citenamefont {Klapwijk}}]{Janssen13}%
  \BibitemOpen
  \bibfield  {author} {\bibinfo {author} {\bibfnamefont {R.~M.~J.}\
  \bibnamefont {Janssen}}, \bibinfo {author} {\bibfnamefont {J.~J.~A.}\
  \bibnamefont {Baselmans}}, \bibinfo {author} {\bibfnamefont {A.}~\bibnamefont
  {Endo}}, \bibinfo {author} {\bibfnamefont {L.}~\bibnamefont {Ferrari}},
  \bibinfo {author} {\bibfnamefont {S.~J.~C.}\ \bibnamefont {Yates}}, \bibinfo
  {author} {\bibfnamefont {A.~M.}\ \bibnamefont {Baryshev}}, \ and\ \bibinfo
  {author} {\bibfnamefont {T.~M.}\ \bibnamefont {Klapwijk}},\ }\href {\doibase
  10.1063/1.4829657} {\bibfield  {journal} {\bibinfo  {journal} {Applied
  Physics Letters}\ }\textbf {\bibinfo {volume} {103}},\ \bibinfo {pages}
  {203503} (\bibinfo {year} {2013})},\ \Eprint
  {http://arxiv.org/abs/https://doi.org/10.1063/1.4829657}
  {https://doi.org/10.1063/1.4829657} \BibitemShut {NoStop}%
\bibitem [{\citenamefont {{van Rantwijk}}\ \emph {et~al.}(2016)\citenamefont
  {{van Rantwijk}}, \citenamefont {{Grim}}, \citenamefont {{van Loon}},
  \citenamefont {{Yates}}, \citenamefont {{Baryshev}},\ and\ \citenamefont
  {{Baselmans}}}]{SpacekidsMUX}%
  \BibitemOpen
  \bibfield  {author} {\bibinfo {author} {\bibfnamefont {J.}~\bibnamefont {{van
  Rantwijk}}}, \bibinfo {author} {\bibfnamefont {M.}~\bibnamefont {{Grim}}},
  \bibinfo {author} {\bibfnamefont {D.}~\bibnamefont {{van Loon}}}, \bibinfo
  {author} {\bibfnamefont {S.}~\bibnamefont {{Yates}}}, \bibinfo {author}
  {\bibfnamefont {A.}~\bibnamefont {{Baryshev}}}, \ and\ \bibinfo {author}
  {\bibfnamefont {J.}~\bibnamefont {{Baselmans}}},\ }\href {\doibase
  10.1109/TMTT.2016.2544303} {\bibfield  {journal} {\bibinfo  {journal} {IEEE
  Transactions on Microwave Theory and Techniques}\ }\textbf {\bibinfo {volume}
  {64}},\ \bibinfo {pages} {1876} (\bibinfo {year} {2016})}\BibitemShut
  {NoStop}%
\bibitem [{\citenamefont {Yates}\ \emph {et~al.}(2017)\citenamefont {Yates},
  \citenamefont {Baryshev}, \citenamefont {Yurduseven}, \citenamefont {Bueno},
  \citenamefont {Davis}, \citenamefont {Ferrari}, \citenamefont {Jellema},
  \citenamefont {Llombart}, \citenamefont {Murugesan}, \citenamefont {Thoen},\
  and\ \citenamefont {Baselmans}}]{Yates2017}%
  \BibitemOpen
  \bibfield  {author} {\bibinfo {author} {\bibfnamefont {S.~J.~C.}\
  \bibnamefont {Yates}}, \bibinfo {author} {\bibfnamefont {A.~M.}\ \bibnamefont
  {Baryshev}}, \bibinfo {author} {\bibfnamefont {O.}~\bibnamefont
  {Yurduseven}}, \bibinfo {author} {\bibfnamefont {J.}~\bibnamefont {Bueno}},
  \bibinfo {author} {\bibfnamefont {K.~K.}\ \bibnamefont {Davis}}, \bibinfo
  {author} {\bibfnamefont {L.}~\bibnamefont {Ferrari}}, \bibinfo {author}
  {\bibfnamefont {W.}~\bibnamefont {Jellema}}, \bibinfo {author} {\bibfnamefont
  {N.}~\bibnamefont {Llombart}}, \bibinfo {author} {\bibfnamefont
  {V.}~\bibnamefont {Murugesan}}, \bibinfo {author} {\bibfnamefont {D.~J.}\
  \bibnamefont {Thoen}}, \ and\ \bibinfo {author} {\bibfnamefont {J.~J.~A.}\
  \bibnamefont {Baselmans}},\ }\href {\doibase 10.1109/TTHZ.2017.2755500}
  {\bibfield  {journal} {\bibinfo  {journal} {IEEE Transactions on Terahertz
  Science and Technology}\ }\textbf {\bibinfo {volume} {7}},\ \bibinfo {pages}
  {789} (\bibinfo {year} {2017})}\BibitemShut {NoStop}%
\bibitem [{\citenamefont {H{\"a}hnle}\ \emph {et~al.}(2018)\citenamefont
  {H{\"a}hnle}, \citenamefont {Bueno}, \citenamefont {Huiting}, \citenamefont
  {Yates},\ and\ \citenamefont {Baselmans}}]{HaehnleCryo}%
  \BibitemOpen
  \bibfield  {author} {\bibinfo {author} {\bibfnamefont {S.}~\bibnamefont
  {H{\"a}hnle}}, \bibinfo {author} {\bibfnamefont {J.}~\bibnamefont {Bueno}},
  \bibinfo {author} {\bibfnamefont {R.}~\bibnamefont {Huiting}}, \bibinfo
  {author} {\bibfnamefont {S.~J.~C.}\ \bibnamefont {Yates}}, \ and\ \bibinfo
  {author} {\bibfnamefont {J.~J.~A.}\ \bibnamefont {Baselmans}},\ }\href
  {\doibase 10.1007/s10909-018-1940-1} {\bibfield  {journal} {\bibinfo
  {journal} {Journal of Low Temperature Physics}\ } (\bibinfo {year} {2018}),\
  10.1007/s10909-018-1940-1}\BibitemShut {NoStop}%
\bibitem [{Top()}]{Toptica}%
  \BibitemOpen
  \href@noop {} {}\bibinfo {note} {TERABEAM 1550 (TOPTICA Photonics AG,
  Lochhamer Schlag 19, 82166 Gräfelfing, Germany)}\BibitemShut {NoStop}%
\bibitem [{\citenamefont {Barends}\ \emph {et~al.}(2010)\citenamefont
  {Barends}, \citenamefont {Vercruyssen}, \citenamefont {Endo}, \citenamefont
  {De~Visser}, \citenamefont {Zijlstra}, \citenamefont {Klapwijk},
  \citenamefont {Diener}, \citenamefont {Yates},\ and\ \citenamefont
  {Baselmans}}]{Barends2010}%
  \BibitemOpen
  \bibfield  {author} {\bibinfo {author} {\bibfnamefont {R.}~\bibnamefont
  {Barends}}, \bibinfo {author} {\bibfnamefont {N.}~\bibnamefont
  {Vercruyssen}}, \bibinfo {author} {\bibfnamefont {A.}~\bibnamefont {Endo}},
  \bibinfo {author} {\bibfnamefont {P.}~\bibnamefont {De~Visser}}, \bibinfo
  {author} {\bibfnamefont {T.}~\bibnamefont {Zijlstra}}, \bibinfo {author}
  {\bibfnamefont {T.}~\bibnamefont {Klapwijk}}, \bibinfo {author}
  {\bibfnamefont {P.}~\bibnamefont {Diener}}, \bibinfo {author} {\bibfnamefont
  {S.}~\bibnamefont {Yates}}, \ and\ \bibinfo {author} {\bibfnamefont
  {J.}~\bibnamefont {Baselmans}},\ }\href {\doibase 10.1063/1.3458705}
  {\bibfield  {journal} {\bibinfo  {journal} {Applied Physics Letters}\
  }\textbf {\bibinfo {volume} {97}},\ \bibinfo {pages} {023508 } (\bibinfo
  {year} {2010})}\BibitemShut {NoStop}%
\end{thebibliography}%


\providecommand{\noopsort}[1]{}\providecommand{\singleletter}[1]{#1}%
\begin{thebibliography}{8}%
\makeatletter
\providecommand \@ifxundefined [1]{%
 \@ifx{#1\undefined}
}%
\providecommand \@ifnum [1]{%
 \ifnum #1\expandafter \@firstoftwo
 \else \expandafter \@secondoftwo
 \fi
}%
\providecommand \@ifx [1]{%
 \ifx #1\expandafter \@firstoftwo
 \else \expandafter \@secondoftwo
 \fi
}%
\providecommand \natexlab [1]{#1}%
\providecommand \enquote  [1]{``#1''}%
\providecommand \bibnamefont  [1]{#1}%
\providecommand \bibfnamefont [1]{#1}%
\providecommand \citenamefont [1]{#1}%
\providecommand \href@noop [0]{\@secondoftwo}%
\providecommand \href [0]{\begingroup \@sanitize@url \@href}%
\providecommand \@href[1]{\@@startlink{#1}\@@href}%
\providecommand \@@href[1]{\endgroup#1\@@endlink}%
\providecommand \@sanitize@url [0]{\catcode `\\12\catcode `\$12\catcode
  `\&12\catcode `\#12\catcode `\^12\catcode `\_12\catcode `\%12\relax}%
\providecommand \@@startlink[1]{}%
\providecommand \@@endlink[0]{}%
\providecommand \url  [0]{\begingroup\@sanitize@url \@url }%
\providecommand \@url [1]{\endgroup\@href {#1}{\urlprefix }}%
\providecommand \urlprefix  [0]{URL }%
\providecommand \Eprint [0]{\href }%
\providecommand \doibase [0]{http://dx.doi.org/}%
\providecommand \selectlanguage [0]{\@gobble}%
\providecommand \bibinfo  [0]{\@secondoftwo}%
\providecommand \bibfield  [0]{\@secondoftwo}%
\providecommand \translation [1]{[#1]}%
\providecommand \BibitemOpen [0]{}%
\providecommand \bibitemStop [0]{}%
\providecommand \bibitemNoStop [0]{.\EOS\space}%
\providecommand \EOS [0]{\spacefactor3000\relax}%
\providecommand \BibitemShut  [1]{\csname bibitem#1\endcsname}%
\let\auto@bib@innerbib\@empty
\bibitem [{\citenamefont {Mattis}\ and\ \citenamefont
  {Bardeen}(1958)}]{MattisBardeen}%
  \BibitemOpen
  \bibfield  {author} {\bibinfo {author} {\bibfnamefont {D.~C.}\ \bibnamefont
  {Mattis}}\ and\ \bibinfo {author} {\bibfnamefont {J.}~\bibnamefont
  {Bardeen}},\ }\href {\doibase 10.1103/PhysRev.111.412} {\bibfield  {journal}
  {\bibinfo  {journal} {Phys. Rev.}\ }\textbf {\bibinfo {volume} {111}},\
  \bibinfo {pages} {412} (\bibinfo {year} {1958})}\BibitemShut {NoStop}%
\bibitem [{\citenamefont {Bardeen}\ \emph {et~al.}(1957)\citenamefont
  {Bardeen}, \citenamefont {Cooper},\ and\ \citenamefont
  {Schrieffer}}]{BCSpaper}%
  \BibitemOpen
  \bibfield  {author} {\bibinfo {author} {\bibfnamefont {J.}~\bibnamefont
  {Bardeen}}, \bibinfo {author} {\bibfnamefont {L.~N.}\ \bibnamefont {Cooper}},
  \ and\ \bibinfo {author} {\bibfnamefont {J.~R.}\ \bibnamefont {Schrieffer}},\
  }\href {\doibase 10.1103/PhysRev.108.1175} {\bibfield  {journal} {\bibinfo
  {journal} {Phys. Rev.}\ }\textbf {\bibinfo {volume} {108}},\ \bibinfo {pages}
  {1175} (\bibinfo {year} {1957})}\BibitemShut {NoStop}%
\bibitem [{\citenamefont {{Kautz}}(1978)}]{Kautz78}%
  \BibitemOpen
  \bibfield  {author} {\bibinfo {author} {\bibfnamefont {R.~L.}\ \bibnamefont
  {{Kautz}}},\ }\href {\doibase 10.1063/1.324387} {\bibfield  {journal}
  {\bibinfo  {journal} {Journal of Applied Physics}\ }\textbf {\bibinfo
  {volume} {49}},\ \bibinfo {pages} {308} (\bibinfo {year} {1978})}\BibitemShut
  {NoStop}%
\bibitem [{\citenamefont {de~Visser}()}]{Pieter14}%
  \BibitemOpen
  \bibfield  {author} {\bibinfo {author} {\bibfnamefont {P.~J.}\ \bibnamefont
  {de~Visser}},\ }\emph {\bibinfo {title} {Quasiparticle dynamics in aluminium
  superconducting microwave resonators}},\ \href@noop {} {Ph.D.
  thesis}\BibitemShut {NoStop}%
\bibitem [{\citenamefont {Collin}(1992)}]{collin_foundations_1992}%
  \BibitemOpen
  \bibfield  {author} {\bibinfo {author} {\bibfnamefont {R.~E.}\ \bibnamefont
  {Collin}},\ }\href {https://books.google.nl/books?id=ywepQwAACAAJ} {\emph
  {\bibinfo {title} {Foundations for {Microwave} {Engineering}}}},\
  {McGraw}-{Hill} series in electrical engineering\ (\bibinfo  {publisher}
  {McGraw-Hill},\ \bibinfo {year} {1992})\BibitemShut {NoStop}%
\bibitem [{\citenamefont {Manual}(2008)}]{manual2008sonnet}%
  \BibitemOpen
  \bibfield  {author} {\bibinfo {author} {\bibfnamefont {E.~U.}\ \bibnamefont
  {Manual}},\ }\href@noop {} {\bibfield  {journal} {\bibinfo  {journal} {Inc.,
  Liverpool, NY}\ } (\bibinfo {year} {2008})}\BibitemShut {NoStop}%
\bibitem [{\citenamefont {{Pascual Laguna}}()}]{Alejandro_CST}%
  \BibitemOpen
  \bibfield  {author} {\bibinfo {author} {\bibfnamefont {A.}~\bibnamefont
  {{Pascual Laguna}}},\ }\href@noop {} {}\bibinfo {howpublished} {private
  communication}\BibitemShut {NoStop}%
\bibitem [{\citenamefont {Pozar}(2011)}]{PozarBook}%
  \BibitemOpen
  \bibfield  {author} {\bibinfo {author} {\bibfnamefont {D.~M.}\ \bibnamefont
  {Pozar}},\ }\href {https://cds.cern.ch/record/882338} {\emph {\bibinfo
  {title} {{Microwave engineering; 4th ed.}}}}\ (\bibinfo  {publisher}
  {Wiley},\ \bibinfo {address} {Hoboken, NJ},\ \bibinfo {year}
  {2011})\BibitemShut {NoStop}%
\end{thebibliography}%
\bibliographystyle{apsrev4-1}

\end{document}


\appendix
\section{Characterization of a superconducting Coplanar Waveguide}
\label{sec:CPW}
The impedance $Z_0$ and effective dielectric constant $\epsilon_{eff}$ of a superconducting Coplanar Waveguide (CPW) can be calculated analytically by evaluating the CPW geometry and the superconductor properties independently, using only the parameters in Table \ref{table}.

These parameters can be easily obtained from experiment with SEM (Scanning Electron Microscope) images of the CPW and DC property measurements of the superconductor. Characterizing the CPW is then split into two parts: First, calculate the surface impedance of the superconductor. Second, calculate the CPW properties using the surface impedance as input.  

\begin{table}\centering
	\begin{tabular}{l|l|l|}
                                    & Parameter & Value                       \\ \hline
\multirow{2}{*}{Superconductor}     & $T_c$     & $14.7\ \mathrm{K}$             \\ \cline{2-3} 
& $\rho_N $ & $102\ \mathrm{\mu \Omega cm}$ \\ \hline
\multirow{3}{*}{CPW geometry}       & $s $      & $1.95\ \mathrm{\mu m}$      \\ \cline{2-3} 
& $w $      & $2.15\ \mathrm{\mu m}$      \\ \cline{2-3} 
& $t $      & $100\ \mathrm{nm}$          \\ \hline
\multirow{2}{*}{Experimental Setup} & $T $      & $0.25\ \mathrm{K}$          \\ \cline{2-3} 
& $f $ & $320..380\ \mathrm{GHz}$    \\ \hline
	\end{tabular}
	\caption{Parameter definitions: $T_c$ is the critical temperature of the superconductor. $\rho_N$ is the superconductors normal state resistivity. The CPW is determined by the line width $s$, the slot width $w$ and the film thickness $t$. Additional dependencies are on the system temperature $T$ and measurement frequency $f$.}
	\label{table}
\end{table}

\subsection{Superconductor}
According to Mattis-Bardeen theory\cite{MattisBardeen}, based on the microscopic BCS theory of superconductivity\cite{BCSpaper}, one can define a conductivity for the superconductor
\begin{equation}
	\sigma = \sigma_1 - i\sigma_2
\end{equation}
analogous with Ohm's law $J = \sigma E$. The real and imaginary parts of the conductivity are given by integrals as  
\begin{equation}
\begin{split}
\frac{\sigma_1}{\sigma_N} & = \frac{2}{\hbar\omega}\int_{\Delta}^{\infty}|f(E)-f(E+\hbar\omega)|g_1(E)dE \\
& + \frac{1}{\hbar\omega}\int_{min(\Delta-\hbar\omega,-\Delta)}^{-\Delta}|1-2f(E+\hbar\omega)|g_1(E)dE 
\end{split}
\end{equation}
\begin{equation}
\frac{\sigma_2}{\sigma_N} = \frac{1}{\hbar\omega}\int_{max(\Delta-\hbar\omega,-\Delta)}^{\Delta}|1-2f(E+\hbar\omega)|g_2(E)dE  
\end{equation}
where $\omega=2\pi f$, $\sigma_N = 1/\rho_N$ is the normal state conductivity, $\Delta \approx 1.76k_BT_c$ is the superconductors gap energy, $f(E)$ is the density of states generally given by the Fermi-Dirac distribution,
\begin{equation}
	f(E) = \frac{1}{1 + \exp(E/k_BT)}
\end{equation}
and $g_1(E)$ and $g_2(E)$ are defined as
\begin{equation}
	g_1(E) = \frac{E^2+\Delta^2+\hbar\omega E}{(E^2-\Delta^2)^{1/2}\left[(E+\hbar\omega)^2-\Delta^2\right]^{1/2}}
\end{equation}
\begin{equation}
g_2(E) = \frac{E^2+\Delta^2+\hbar\omega E}{(\Delta^2-E^2)^{1/2}\left[(E+\hbar\omega)^2-\Delta^2\right]^{1/2}}.
\end{equation}

The second integral in $\sigma_1$ describes the pair-breaking process of Cooper pairs when $\hbar\omega >= 2\Delta$ and is zero for frequencies below the gap frequency. 

\begin{figure}[ht]
	\centering
	\includegraphics[width=0.5\textwidth]{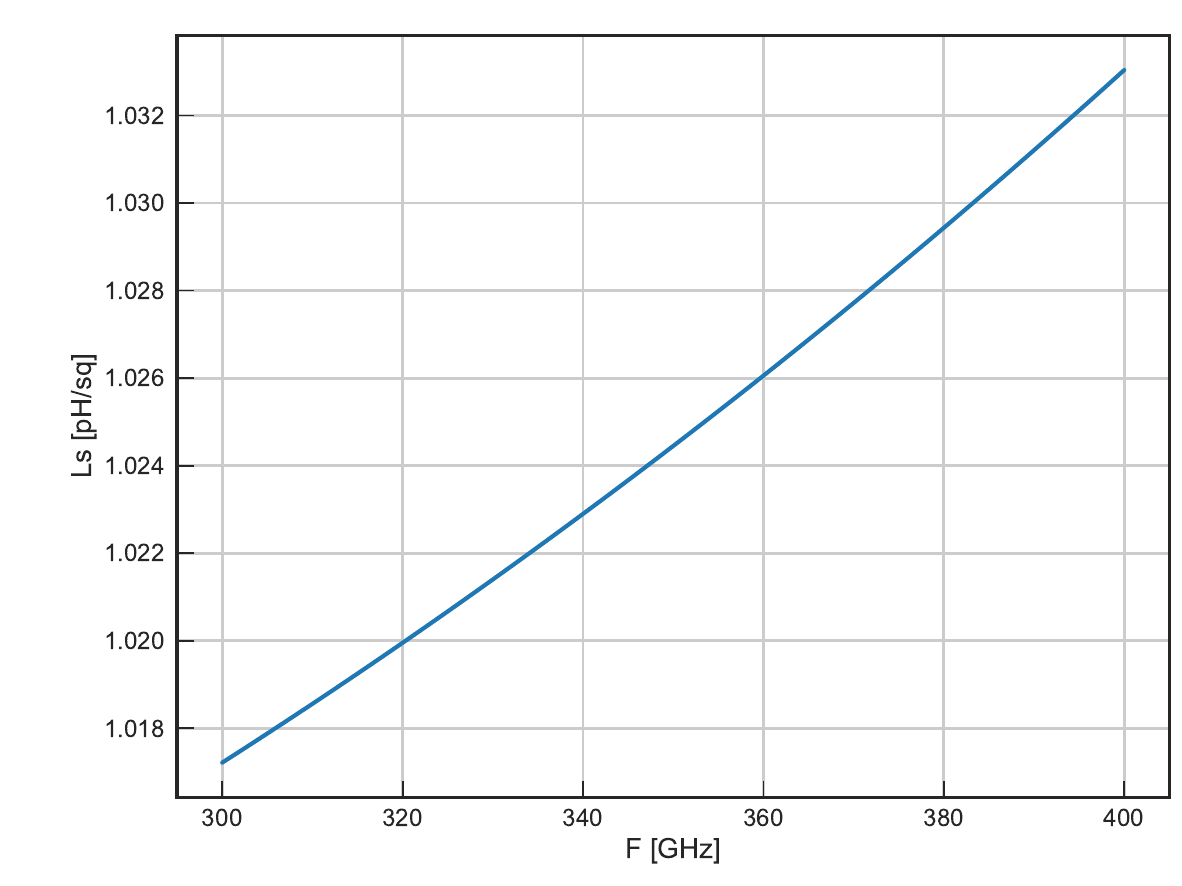}
	\caption{Frequency dependent surface inductance calculated using the values in table \ref{table}}
	\label{fig:Ls} 
\end{figure}

From the complex conductivity, the surface impedance then follows from \cite{Kautz78} as 
\begin{equation}\label{Zs}
	Z_s = \sqrt{\frac{i\mu_0\omega}{\sigma}}\coth(t\sqrt{i\omega\mu_0\sigma}) = R_s + i\omega L_s
\end{equation}
given in $[Z_s] = \Omega/\square$ with the surface resistance $R_s$ and the surface inductance
\begin{equation}\label{Ls}
L_s = \frac{\operatorname{Im}(Z_s)}{\omega}.
\end{equation}
This surface inductance is due to the acceleration of cooper pairs in an AC field while the surface resistance is dependent on the number of unpaired electrons (quasiparticles) in the superconductor. The resulting $L_s$, which is shown exemplary in Fig. \ref{fig:Ls} for the values given in Table \ref{table}, has a frequency dependence which is negligible for $\hbar\omega << \Delta$, but becomes significant for $\hbar\omega \gtrsim  \Delta/3$. While various simplifications for the calculation of $L_s$ are possible, these usually break down in the limit of thin films and/or high frequencies and generally do not reproduce this frequency dependence.

\subsection{Superconducting CPW}
In a superconducting CPW, an effective dielectric constant can be determined from its phase velocity $v_{ph}$ as
\begin{equation}\label{epseff1}
\epsilon_{eff}= \frac{c^2}{v_{ph}^2}= c^2(L_g+L_k)C_l.
\end{equation}
with the line capacitance $C_l$, the geometric inductance $L_g$ and the kinetic inductance $L_k$, all given per unit length. Equivalently, the characteristic impedance is given by 
\begin{equation}
	Z_0= \sqrt{\frac{L_g+L_k}{C_l}}
\end{equation} 
The line capacitance and geometric inductance are determined by the dimensions of the CPW and are given by
\begin{equation}
L_g = \frac{\mu_0K(k')}{4K(k)}
\end{equation}
\begin{equation}
C_l = 4\epsilon_0\epsilon_{eff,geo}\frac{K(k)}{K(k')}
\end{equation}
where $k = s/(s+2w)$, $k'^2 = 1 - k^2$, $K$ is the complete elliptic integral of the first kind and $\epsilon_{eff,geo}\approx(1+\epsilon_{subs})/2$ is the effective dielectric constant visible to the line capacitance due to the substrate below the CPW line. 

The kinetic inductance $L_k$ is dependent on both the superconductor properties and the CPW geometry and can be calculated analytically as 
\begin{equation}
	L_k = g_cL_{s,c}+g_gL_{s,g}
\end{equation}
where $L_{s,c}$ and $L_{s,g}$ are the surface inductances of the central line and groundplane as given by eqs.\ref{Zs} and \ref{Ls}, and $g_c$ and $g_g$ their respective geometry factors \cite{Pieter14}
\begin{equation}
	g_c = \frac{1}{4s(1-k^2)K^2(k)}\left[\pi + \ln\left(\frac{4\pi s}{t}\right) - k\ln\left(\frac{1+k}{1-k}\right)\right]
\end{equation} 
\begin{equation}
g_g = \frac{1}{4s(1-k^2)K^2(k)}\left[\pi + \ln\left(\frac{4\pi (s+2w)}{t}\right) - \frac{1}{k}\ln\left(\frac{1+k}{1-k}\right)\right].
\end{equation}
The geometry factors were originally described in \textit{Foundations for Microwave Engineering} by R.E.Collin\cite{collin_foundations_1992} for the losses in the central line and groundplane of a CPW, and have been intuitively adapted to apply for the kinetic inductance contribution. This adaptation has been verified by comparison with simulations, e.g. Sonnet, and has been found to be in excellent agreement with experimental results over the last years.
In the case of a single film CPW with surface inductance $L_s = L_{s,c} = L_{s,g}$, eq. \ref{epseff1} can then be rewritten as
\begin{equation}
	\epsilon_{eff} = c^2(L_g + gL_s)C_l
\end{equation}
with $g = g_c+g_g$. 

\begin{figure}[ht]
	\centering
	\includegraphics[width=0.5\textwidth]{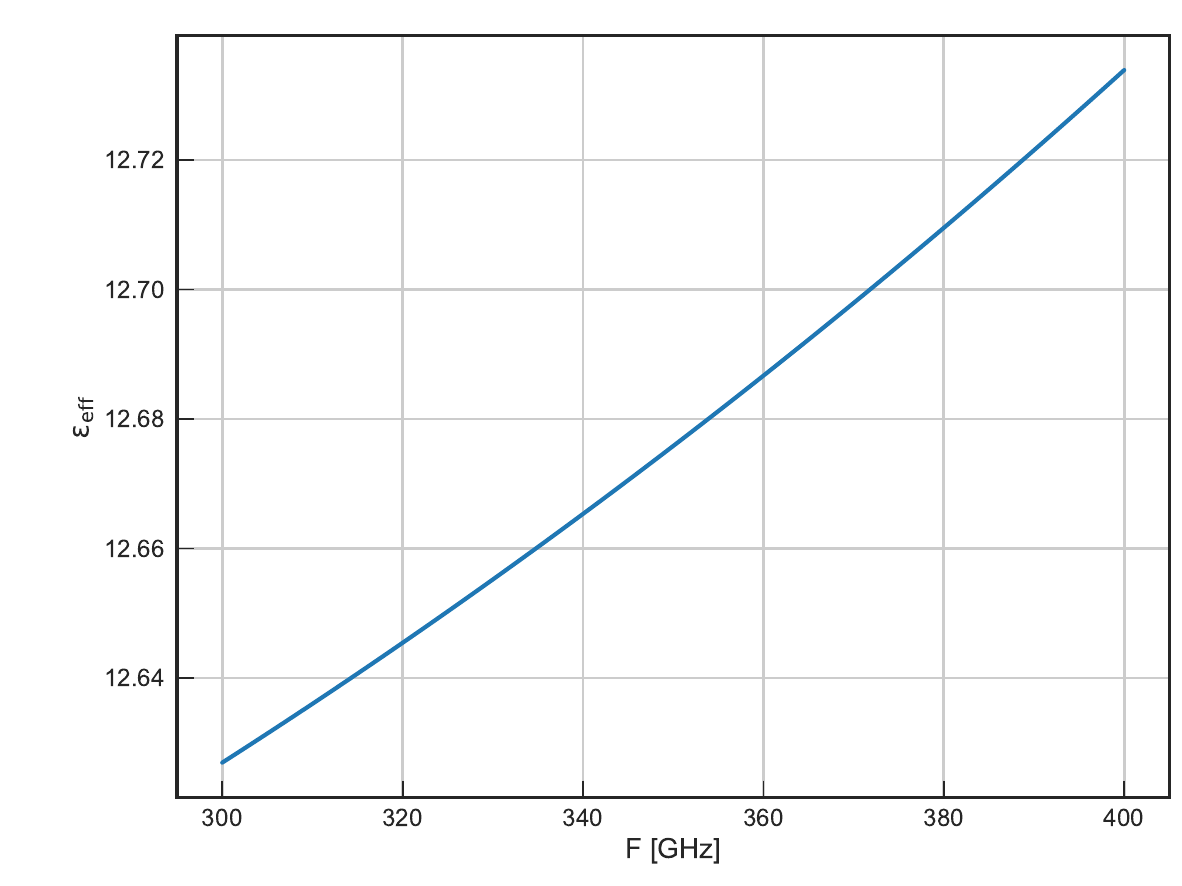}
	\caption{Frequency dependent dielectric constant calculated using the values in table \ref{table}}
	\label{fig:epseff} 
\end{figure}

Figure \ref{fig:epseff} shows the effective dielectric constant as a function of frequency for the values given in Table \ref{table}, where a large kinetic inductance is achieved by choosing a superconductor with large $L_s$, corresponding to a high resistivity and thin film, and designing a narrow CPW with large $g$. In this configuration, the $\epsilon_{eff}$ is dominated by the kinetic inductance, showing the same frequency dependence.

Finally, the internal quality factor of a superconducting CPW due to ohmic losses from quasiparticles is given by 
\begin{equation}
	Q_i = \frac{1}{\alpha_k}\frac{\omega L_s}{R_s}
\end{equation}
with the kinetic inductance fraction $\alpha_k = L_k/(L_g+L_k)$. For the given example values, we obtain $Q_i \approx 10^{49}$, which is negligible compared to other loss sources.

\section{Simulating an on-chip Fabry-P{\'e}rot Resonator}

The Fabry-P{\'e}rot resonator (FP) is very long compared to the relevant wavelengths ($L_{FP}\gtrsim40\lambda$) with feature sizes much smaller than a wavelength, due to the narrow line. This makes a simulation of the full structure impractical due to the required fine mesh and large box size. It is therefore preferable to split the resonator into its separate components, i.e. the coupling structures and the transmission line, and cascade them using ABCD matrices. Three simulation setups to determine the coupling strength as well as the radiation loss of both transmission line and coupler will be shown here. All simulations are performed in Sonnet\cite{manual2008sonnet}, which is a commercial 3D planar EM software capable of simulating superconducting structures at high frequencies.

\subsection{Coupler Simulation}
\label{sec:coupler}
\begin{figure}[ht]
	\centering
	\includegraphics{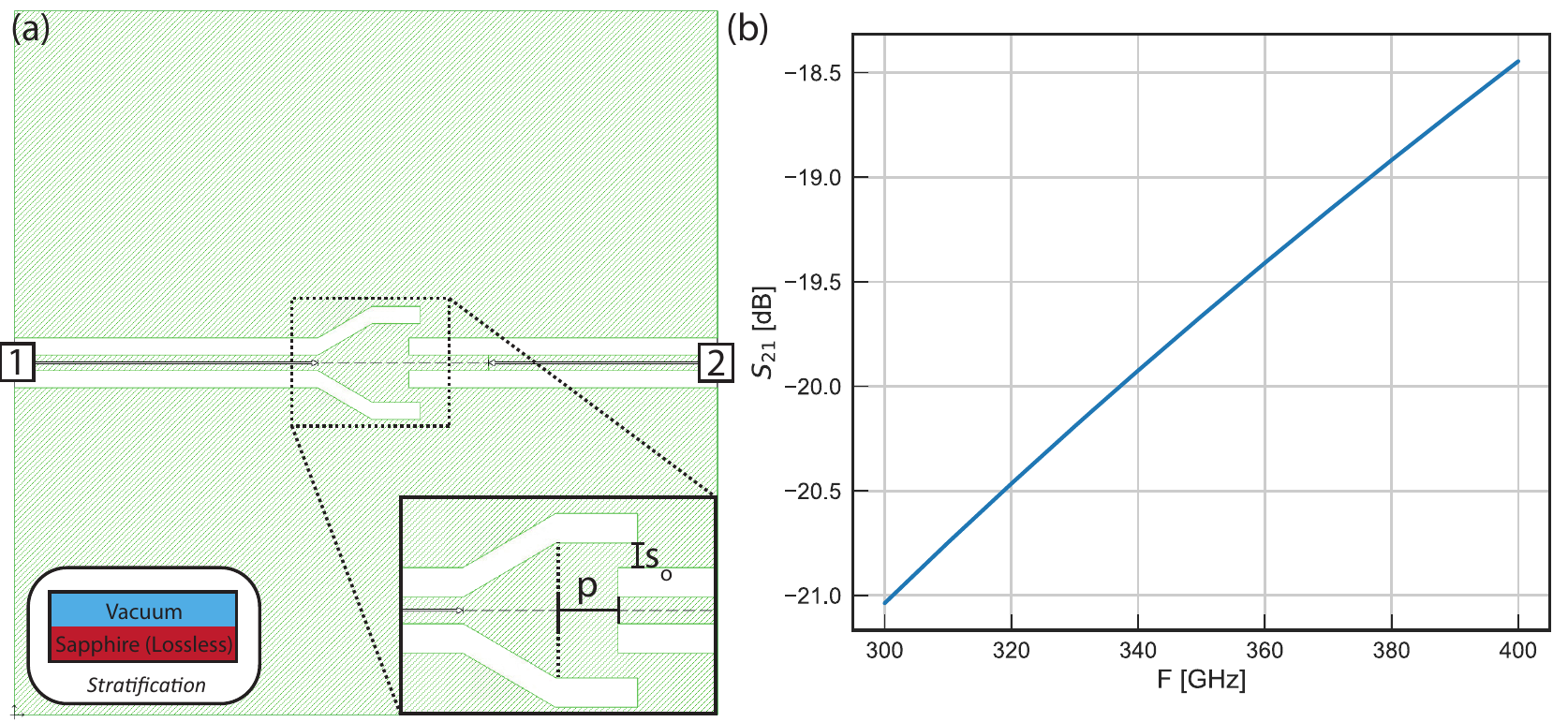}
	\caption{(a) Sonnet box of the coupler simulation with inset showing the stratification and a zoom-in on the the coupling structure. NbTiN is shown in green; substrate in white. (b)  Simulated $S_{21}$ of the measured chip, where $s = 1.95\ \mathrm{\mu m}$ and $w = 2.15\ \mathrm{\mu m}$. Output from Sonnet, with $50 \Omega$ port impedance.}
	\label{fig:coupler} 
\end{figure}

The couper is simulated in a small box of $32\times32\ \mathrm{\mu m}^2$ as shown in Fig. \ref{fig:coupler}a), resulting in the scattering matrix $S$, with $S_{21}$ shown in Fig. \ref{fig:coupler}b) as function of frequency. The box size is chosen as small as possible without impacting the simulation result. Both top and bottom of the box are defined as lossless metal, with a layer of vacuum above and a layer of lossless Sapphire below the metalization with $h_{diel}=500\ \mathrm{\mu m}$ for both layers. Superconducting NbTiN is implemented using the "general model" with zero resistance and a finite surface impedance $L_s$. A lookup table with the frequency dependence for $L_s$, generated using Appendix \ref{sec:CPW}, is used as input for the model. 

The coupling structure is implemented as an overlapping coupler with a short to ground, where the coupling strength can be tuned by changing the overlap parameter $p$. Port 1 on the left side is connecting to either the antenna or the detector, while port 2 is part of the FP resonator. The position of the overlapping slots is adjusted for different CPW line widths on the resonator side to keep $s_{0}=2\ \mathrm{\mu m}$ constant. Due to the frequency dependence of $L_s$ and consequently the line impedance $Z_0$, the line will be slightly mismatched with the constant port impedance $Z_p$. To account for this, a reference plane close to the coupler is used and the port impedances are set to $50\ \Omega$. The scattering matrix for the correct port impedance is then retrieved in post processing. This also makes optimization in the design phase easier, as the port impedance does not need to be adjusted in the simulation setup if the line impedance changes.

In order to obtain the coupler geometry for a desired $Q_c$ of the FP resonator, a sweep of $p$ is performed from $300$ to $400\ \mathrm{GHz}$, using a cell size of $0.5\ \mathrm{\mu m}$ with the smallest feature size of $2\ \mathrm{\mu m}$. The simulations for the fabricated chips are carried out with the measured dimensions and a cell size of $0.05\ \mathrm{\mu m}$ to sample the overetched geometry.

\subsection{Radiation Loss Simulation} 
\subsubsection{CPW line loss} \label{sec:lineloss}
\begin{figure}[ht]
	\centering
	\includegraphics{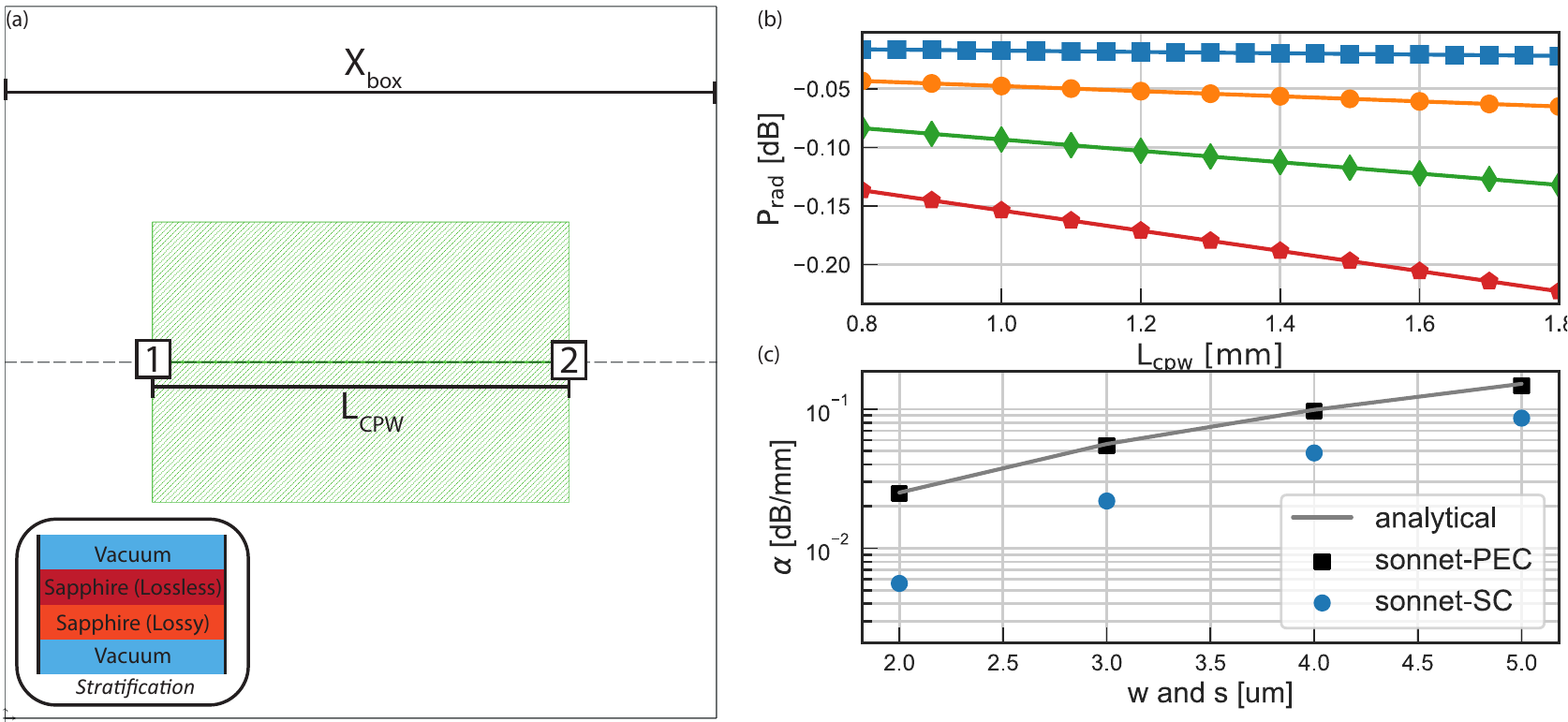}
	\caption{(a) Sonnet box to simulate the radiation loss of a straight CPW line with inset showing the stratification of the dielectric layers. (b) Simulated loss as function of line length for different NbTiN CPWs where s=w ($2\ \mathrm{\mu m}$: \textit{blue}; $3\ \mathrm{\mu m}$: \textit{orange}; $4\ \mathrm{\mu m}$: \textit{green}; $5\ \mathrm{\mu m}$: \textit{red}). The Straight lines are linear fits. (c) Loss factor alpha as function of CPW line width. Sonnet-SC corresponds to the fitted values of plot (b). Sonnet-PEC is compared to the analytical solution for a PEC CPW, showing good agreement.}
	\label{fig:lineloss} 
\end{figure}

Retrieving radiation loss of any structure in Sonnet requires a careful setup of the simulation. Any loss in the structure corresponds to power that does not reach the ports of the simulation and can therefore be obtained from the scattering matrix as
\begin{equation}
	P_{rad} = 1 - P_{out} = 1 - (|S_{11}|^2+|S_{21}|^2).
\end{equation}
The Sonnet box used to simulate the radiation loss in a CPW, shown in Fig. \ref{fig:lineloss}, fulfills the following requirements for an accurate result:
\begin{itemize}
	\item The structure is able to radiate freely.
	\item Radiated power is not absorbed in the ports.
	\item The structure is otherwise lossless (e.g. ohmic losses, dielectric losses).
\end{itemize}

The CPW can radiate when the propagating field can couple to modes in the surrounding media, i.e. the vacuum above and the Sapphire substrate below. However, the lossless metal walls of the Sonnet box act as a waveguide with a cut-off frequency given by the box size, below which no modes can be excited. Therefore, the box is made sufficiently large compared to the freespace wavelength $\lambda_0$ ($Y_{box} = X_{box}\gtrsim 2\lambda_0 = 2048\ \mathrm{\mu m}$) to allow all relevant modes to be excited. All relevant dimensions used in the simulation are also summarized in table \ref{tab:lineloss}

\begin{table}[]
	\label{tab:lineloss}
	\begin{tabular}{l|llll}
		Box size $\mathrm{[\mu m^2]}$                   & $2048\times2048$ &  &  &  \\ \cline{1-2}
		Cell size $\mathrm{[\mu m^2]}$                  & $0.5\times0.5$   &  &  &  \\ \cline{1-2}
		Groundplane width $\mathrm{[\mu m]}$            & 400              &  &  &  \\ \cline{1-2}
		CPW length $\mathrm{[\mu m]}$                   & 800..1800        &  &  &  \\ \cline{1-2}
		Vacuum layer thickness                          & $\lambda_0/4$    &  &  &  \\ \cline{1-2}
		Lossless substrate thickness $\mathrm{[\mu m]}$ & 100              &  &  &  \\ \cline{1-2}
		Lossy substrate thickness $\mathrm{[\mu m]}$    & 100,000          &  &  &  \\ \cline{1-2}
		Loss tangent                                    & 1                &  &  &  \\ \cline{1-2}
		F [GHz]											& 350				& & & 
	\end{tabular}
	\caption{Parameters used for the radiation loss simulation.}
\end{table}

The simulation is set up as shown in Fig. \ref{fig:lineloss}a) with superconducting NbTiN and a $100\ \mathrm{\mu m}$ thick lossless Sapphire substrate equivalent to section \ref{sec:coupler}, resulting in no ohmic or dielectric losses. However, for a fully enclosed lossless metal box with no lossy components, any radiated power will be reflected back and eventually absorbed in one of the ports and result in $P_{rad}=0$. To avoid this in the Sonnet simulation, the top and bottom walls are set to free space, where radiation at the boundary is absorbed. However, surface waves in the Sapphire substrate are still confined by reflections at the sidewall. A thick lossy layer of Sapphire is placed below the lossless layer in order to attenuate these surface waves, before they can be absorbed in the ports. Ports are set up as co-calibrated internal ports in a push-pull configuration with a floating ground connection (see Fig. \ref{fig:lineloss}a)) and without de-embedding. To avoid reflections, the port impedance is set to the line impedance, which is retrieved from a separate simulation. The metalization is confined to a patch in the middle of the box such that the sidewalls do not affect the radiating structure.
Finally, the vacuum layer above the CPW needs to be precisely $\lambda_0/2$ thick, otherwise numerical issues arise in the simulation.

In summary, the following design rules need to followed:
\begin{itemize}
	\item A large box size compared to the wavelength $Y_{box} = X_{box}\gtrsim 2\lambda_0$.
	\item The top and bottom wall of the box are free space.
	\item A precise vacuum layer thickness above the metalization $t_{vac} = \lambda_0/2$.
	\item Two layers of substrate, one lossless layer ($\tan\delta=0$) of medium thickness directly below the metalization and one lossy, very thick layer ($\tan\delta=1$) below that.
	\item Co-calibrated internal ports with a floating ground connection.
	\item The superconductor is a metal using the general model with $L_s$ set as the surface inductance in $\mathrm{pH/\square}$.
\end{itemize}

To make certain that the simulated $P_{rad}$ is not affected by any other systematic errors in the simulation, a sweep of the CPW length $L_{cpw}$ is performed and a linear fit through the resulting $P_{rad}(L_{cpw})$ is used to determine the loss factor $\alpha$, as shown in Fig. \ref{fig:lineloss}b). The groundplane can be kept at a constant width for this purpose. The corresponding internal quality factor is then given by $Q_i = \beta/(2\alpha)$, with $Q_i = 15100$ for the narrowest line with $s=w=2\ \mathrm{\mu m}$. 

This simulation method was verified by comparing a PEC simulation ($L_s = 0\ \mathrm{pH/\square}$) with analytical models, which found good agreement as shown in Fig. \ref{fig:lineloss}c). Additionally, good agreement was found with simulations of superconducting CPW carried out in the 3D EM software CST\cite{Alejandro_CST}.

\subsubsection{Resonator loss}
\begin{figure}[ht]
	\centering
	\includegraphics[width=\textwidth]{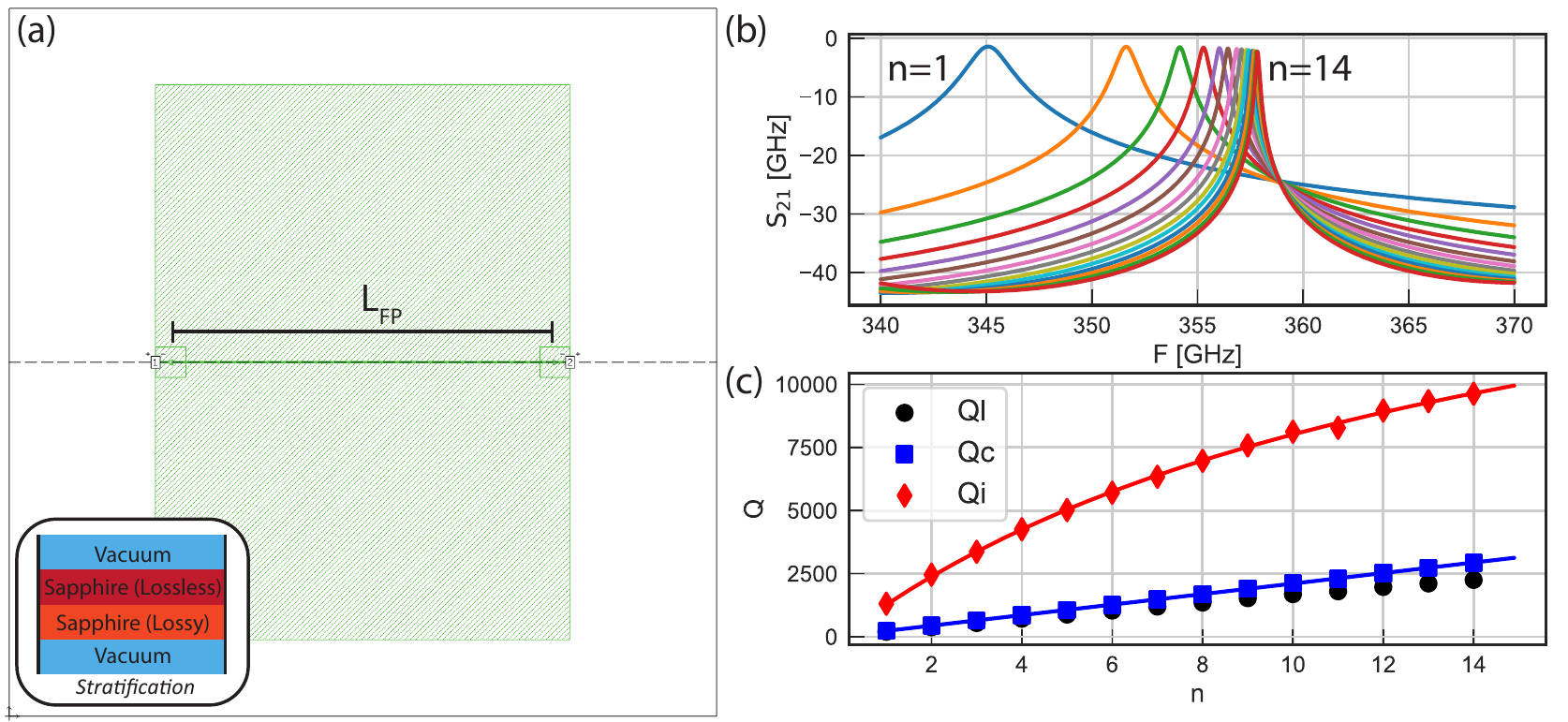}
	\caption{(a) Sonnet box to simulate the radiation loss of a short Fabry-P{\'e}rot resonator, with inset showing the stratification of the dielectric layers. (b) Simulated $S_{21}$ for different resonator lengths, corresponding to different mode numbers $n$. (c) Quality factors extracted from the peaks in figure b shown as points with fits for $Q_i$ and $Q_c$ shown as lines. }
	\label{fig:resonatorloss} 
\end{figure}

While a full length FP resonator is too large to implement in Sonnet, shorter resonators can be simulated as shown in Fig \ref{fig:resonator}a) to obtain the loss at a given frequency as function of mode number. For this purpose, the same box as in section \ref{sec:lineloss} is used, but the simple straight line is exchanged for a 2-port FP resonator with the couplers as designed in \ref{sec:coupler} and a resonator length such that 
\begin{equation}
	L_{res} = n \frac{c}{2F\sqrt{\epsilon_{eff}}}
\end{equation}
with the resonance frequency $F=350\ \mathrm{GHz}$, the dielectric constant of the line $\epsilon_{eff}$ and the mode number $n$, which is varied from 1 to 14. The simulation for each mode number is carried out in a small range around the resonance frequency, resulting in a $S_{21}$ peak which is dependent on mode number as shown in Fig. \ref{fig:resonatorloss}b) for a CPW with $s=w=2\ \mathrm{\mu m}$. The downshift in resonance frequency at small mode numbers, shown in Fig. \ref{fig:resonatorloss}b), is due to the coupling inductance, which represents a larger fraction of the total resonator inductance for shorter resonators.
The loaded quality factor $Q_L$ of the peak is given by
\begin{equation}
	\frac{1}{Q_L} = \frac{1}{Q_i} + \frac{1}{Q_c} = \frac{1}{Q_{i,l}} + \frac{1}{Q_{i,c}}+\frac{1}{Q_c}
\end{equation}
with the coupling strength $Q_c$, the internal loss of the line $Q_{i,l}$ and the loss at the coupler $Q_{i,c}$. 
Both $Q_{i,c}$ and $Q_c$ are linear in mode number and can be expressed as $Q_{i,c} = nQ_{i,c1}$ and $Q_c = nQ_{c1}$, where the index 1 corresponds to the value at $n=1$. This can be intuitively understood as a reduced impact of the coupler on the resonator behaviour when the resonator becomes longer than the wavelength. 
The peak height of the resonator is given by 
\begin{equation}
	|S_{21}^{max}| = \frac{Q_L}{Q_c},
\end{equation}
and can therefore be used to distinguish between internal losses and the coupling strength. Fitting $Q_c$ and $Q_i$ results in $Q_{c1}=212$, $Q_{i,c1}=1307$ and $Q_{i,l}=16980$. As $Q_{i,l}$ is independent of mode number, the resonator is in a $Q_c$ limited regime for low $n$ and transitions to a $Q_{i,l}$ dominated regime at high $n$, while $Q_{i,c}$ is negligible in both regimes.
The obtained value for $Q_{i,l}$ shows good agreement with the pure line simulation of $Q_i = 15100$ in section \ref{sec:lineloss}. Slight deviations between these two values are expected, as a resonating structure has a different current distributions compared to a simple straight line, thus affecting the radiating fields.

\subsection{ABCD-Matrix method}
\begin{figure}[ht]
	\centering
	\includegraphics[width=\textwidth]{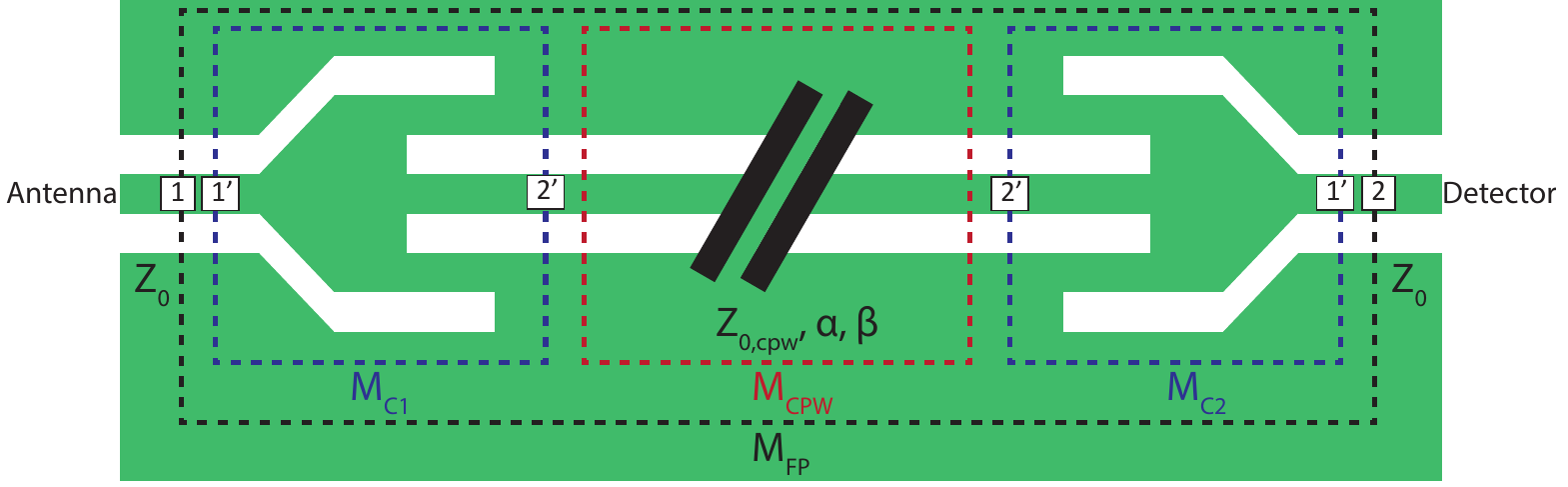}
	\caption{Schematic of a Fabry-P{\'e}rot resonator split into its individual components with associated ABCD matrices.}
	\label{fig:resonator} 
\end{figure}

The transmission $S_{21}$ through the Fabry-P{\'e}rot resonator can be obtained by splitting it into separate network elements, solving their individual behaviour, and then cascading the resulting ABCD matrices \cite{PozarBook}
\begin{equation} \label{eq:Mfp}
	M_{FP} = M_{C1}M_{CPW}M_{C2} = 
	\begin{bmatrix}
	A & B\\
	C & D
	\end{bmatrix}
\end{equation} 
where $M_{FP}$ is the ABCD matrix of the full resonator and $M_{C1}$, $M_{CPW}$, $M_{C2}$ correspond to the couplers and CPW line as shown in Fig. \ref{fig:resonator}. This approach requires the use of ABCD matrices, as scattering matrices can not be cascaded in this way. 

The individual ABCD matrices are given as:
\begin{equation}
M_{C1} = M_{C2} = 
\begin{bmatrix}
\frac{(1+S_{1'1'})(1-S_{2'2'})+S_{1'2'}S_{2'1'}}{2S_{2'1'}} &
Z_p\frac{(1+S_{1'1'})(1+S_{2'2'})-S_{1'2'}S_{2'1'}}{2S_{2'1'}} \\
\frac{1}{Z_p}\frac{(1-S_{1'1'})(1-S_{2'2'})-S_{1'2'}S_{2'1'}}{2S_{2'1'}} &
\frac{(1-S_{1'1'})(1+S_{2'2'})+S_{1'2'}S_{2'1'}}{2S_{2'1'}}
\end{bmatrix}
\end{equation}
where $S_{ij}$ and $Z_p$ are the scattering parameters and the port impedance of the coupler simulation as given in section \ref{sec:coupler}, and  
\begin{equation}
M_{CPW} =
\begin{bmatrix}
\cosh(\gamma L_{FP}) & Z_{0,cpw}\sinh(\gamma L_{FP}) \\
\frac{1}{Z_{0,cpw}}\sinh(\gamma L_{FP}) & \cosh(\gamma L_{FP})
\end{bmatrix}.
\end{equation}
where $Z_{0,cpw}$ is the characteristic impedance of the FP line, $L_{FP}$ is the resonator length and  $\gamma = \alpha + i\beta$ is the complex propagation constant with the loss factor $\alpha$ and the propagation constant $\beta$. The loss factor $\alpha$ can be obtained from the simulation of section \ref{sec:lineloss}, while $Z_{0,cpw}$ and $\beta$ can be obtained either from simulation or analytically as described in section \ref{sec:CPW}.

\begin{figure}[ht]
	\centering
	\includegraphics[width=0.5\textwidth]{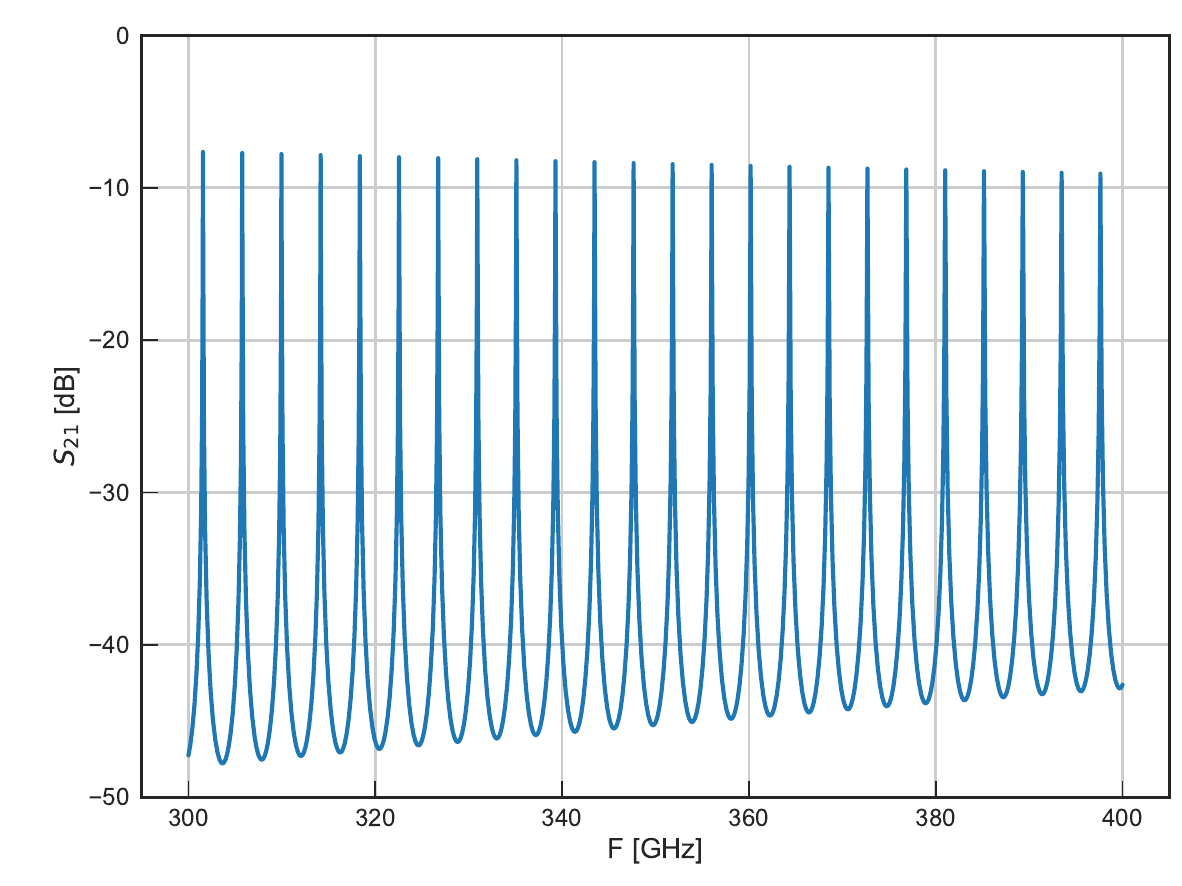}
	\caption{Calculated Fabry-P{\'e}rot transmission based on Sonnet simulations of the coupling strength and radiation loss for a CPW with $s=1.95\ \mathrm{\mu m}$ and $w=2.15\ \mathrm{\mu m}$.}
	\label{fig:resonator_spectrum} 
\end{figure}

The transmission $S_{21}$ through the resonator can then be retrieved from eq. \ref{eq:Mfp}
\begin{equation}
	S_{21} = \frac{2(AD-BC)}{A+B/Z_0+CZ_0+D}
\end{equation}
where $Z_0$ is the characteristic impedance of the transmission line outside the resonator, assuming the lines to the antenna and the detector are identical. 
The resulting spectrum using the example simulations from sections \ref{sec:coupler} and \ref{sec:lineloss} with a resonator length of $L_{FP} = 10\ \mathrm{mm}$ is shown in Fig.\ref{fig:resonator_spectrum}, clearly showing the characteristic peaks of the Fabry-P{\'e}rot, with a reduced peak height due to the lossy CPW.

\bibliography{Bibliography}
\bibliographystyle{apsrev4-1}